\documentclass[11pt]{article}
\usepackage{epsfig}
\newenvironment{proof}{\noindent {\bf Proof.}}{\bigskip}

\usepackage{epsfig}
\usepackage{amsmath}
\usepackage{amssymb}
\usepackage{color}

{ %
  \newtheorem{lemma}{\textbf{Lemma}}[section]%
  \newtheorem{theorem}[lemma]{\textbf{Theorem}}%
  \newtheorem{proposition}{Proposition}%
  {%
    }%
  {%
    }%
  {%
    }%
}

\begin{document}

\centerline{\Large\bf Shortest path problem in rectangular complexes}

\medskip
\centerline{\Large\bf  of global  nonpositive curvature}

\vspace{10mm}

\centerline{\large {\sc Victor Chepoi and Daniela Maftuleac}}

\vspace{5mm}

\begin{center}
Laboratoire d'Informatique Fondamentale de Marseille,\\[0.1cm]
Universit\'e de la M\'editerran\'ee, Facult\'e des Sciences de\\[0.1cm]
Luminy, F-13288 Marseille Cedex 9, France, \\[0.1cm]
{\em\{chepoi,maftuleac\}@lif.univ-mrs.fr}
\end{center}


\noindent

\vspace{7mm}
\begin{footnotesize} \noindent {\bf Abstract.} CAT(0) metric spaces constitute  a far-reaching common generalization of Euclidean and hyperbolic spaces and simple polygons: any two points $x$ and $y$ of a CAT(0) metric space are connected by a unique shortest path $\gamma(x,y).$ In this paper, we present an efficient algorithm for answering two-point distance queries in CAT(0) rectangular complexes and two of theirs subclasses, ramified rectilinear polygons (CAT(0) rectangular complexes in which the links of all vertices are bipartite graphs) and squaregraphs (CAT(0) rectangular complexes arising from plane quadrangulations in which all inner vertices have degrees $\ge 4$). Namely, we show that for a CAT(0) rectangular complex $\mathcal K$ with $n$ vertices, one can construct a data structure $\mathcal D$ of size $O(n^2)$ so that, given any two points $x,y\in \mathcal K,$ the shortest path $\gamma (x,y)$ between $x$ and $y$ can be computed in $O(d(p,q))$ time, where $p$ and $q$ are  vertices of two faces of $\mathcal K$ containing the points $x$ and $y,$ respectively, such that $\gamma(x,y)\subset {\mathcal K}(I(p,q))$ and $d(p,q)$ is the distance between $p$ and $q$ in the underlying  graph of $\mathcal K$. If $\mathcal K$ is a ramified rectilinear polygon, then one can construct a data structure $\mathcal D$ of optimal size $O(n)$ and answer two-point shortest path queries in $O(d(p,q)\log\Delta)$ time, where $\Delta$ is the maximal degree of a vertex of $G(\mathcal K).$ Finally, if $\mathcal K$ is a squaregraph, then one can construct a data structure $\mathcal D$ of  size $O(n\log n)$ and answer two-point shortest path queries in $O(d(p,q))$ time.
\end{footnotesize}

\bigskip
\noindent {\it Keywords.} Shortest path problem, rectangular complex, geodesic $l_2$-distance, global nonpositive curvature.

\section{Introduction}
The shortest path problem is one of the best-known algorithmic problems with many applications in routing, robotics, operations research,
motion planning, urban transportation, and terrain navigation. This fundamental problem was intensively studied both in discrete settings like graphs and networks (see, e.g.,
Ahuja, Magnanti, and Orlin \cite{AhMaOr}) as well as in geometric spaces (simple polygons, polygonal domains with obstacles,
polyhedral surfaces, terrains; see, e.g., Mitchell \cite{Mi}). In the case of graphs $G=(V,E)$ in which all edges have non-negative lengths, a well-known
algorithm of Dijkstra 
allows us to compute a tree of shortest paths from any source vertex to all other vertices of the graph. In simple polygons $P$ endowed with the (intrinsic) geodesic distance, each pair of points $p,q\in P$ {\it can be connected by a unique shortest path.}
Several algorithms for computing shortest paths inside a simple polygon are known in the literature \cite{GuiHer, GuiHerLeShaTar, HerSu, LeePre,ReSto}, and all are based on a triangulation of  $P$ in a preprocessing step (which can be done in linear time due to Chazelle's algorithm \cite{Cha}). The algorithm of Lee and Preparata \cite{LeePre}  finds the shortest path between two points of a triangulated simple polygon in linear time ({\it two-point shortest path queries}). Given a source point, the algorithm of Reif and Storer \cite{ReSto} produces in $O(n\log n)$ time a search structure (in the form of a shortest path tree) so that the shortest path from any query point to the source can be found in time linear in the number of edges of this path (the so-called {\it single-source shortest path queries}).  Guibas et al. \cite{GuiHerLeShaTar}  return a similar search
structure, however their preprocessing step takes only linear time once the polygon is triangulated (see Hersberger and Snoeyink \cite{HerSno} for a significant simplification of the original algorithm of \cite{GuiHerLeShaTar}). Finally, Guibas and Hersberger \cite{GuiHer} showed how to preprocess a triangulated simple polygon $P$ in linear time to support shortest-path queries between any two points $p,q\in P$ in time proportional to the number of edges of the shortest path between $p$ and $q.$ Note that the last three mentioned algorithms also return in $O(\log n)$ time the distance
between the queried points. In the case of shortest path queries in general polygonal domains $D$  with holes, the simplest approach is to compute at the preprocessing step the visibility graph
of $D.$ Now, given two query points $p,q,$ to find a shortest path between $p$ and $q$ in $D$ (this path is no longer unique), it suffices to compute this path in the visibility graph of $D$ augmented with two vertices $p$ and $q$ and all edges corresponding to vertices of $D$ visible from $p$ or $q;$ for a detailed description of how to efficiently construct the visibility graph, see the survey \cite{Mi} and the book \cite{BeChKrOv}.  An alternative paradigm is the so-called {\it continuous Dijkstra} method, which was first applied  to the shortest path problem in general polygonal domains by Mitchell \cite{Mi1} and subsequently improved to a nearly optimal algorithm by Hershberger and Suri \cite{HerSu}; for an extensive overview of this method and related references, see again
the survey by Mitchell \cite{Mi}.

In this paper, we present an algorithm for efficiently solving two-point shortest path queries  in CAT(0) rectangular complexes, i.e., rectangular complexes of global non-positive curvature. CAT(0) metric spaces have been introduced by M. Gromov in his seminal paper \cite{Gr} and investigated in many recent mathematical papers; in particular,
CAT(0) spaces  play a vital role in geometric group
theory \cite{ChDrHa,Ni,Ro,Sa} (CAT(0) cubical complexes also occur in reconfigurable systems \cite{GhLV,GhPe} and metric graph theory \cite{BaCh_survey}); 
CAT(0) metric spaces  can be characterized  as the geodesic metric spaces in which
any two points can be joined by a unique geodesic shortest path, therefore they represent a far-reaching generalization of geodesic metrics in  simple polygons. Several papers are devoted to algorithmic problems in particular CAT(0) spaces. For example, the recent paper by Fletcher et al. \cite{FlMoPhVen} investigates algorithmic questions related to computing approximate convex hulls and centerpoints  of point-sets in the CAT(0) metric space $P(n)$ of all positive definite $n\times n$ matrices. Billera et al.  \cite{BiHoVo} showed that the space of all phylogenetic  trees
 defined on the same set of leaves can be viewed as a CAT(0) cubical complex. Subsequently, the question of whether the distance and the shortest path between two trees in this CAT(0) space can be computed
in polynomial (in the number of leaves) time was raised. Recently, Owen and Provan \cite{OwPro} solved this question in the affirmative; the paper \cite{ChaHo} reports on the implementation of the algorithm of \cite{OwPro}.

The remaining part of the paper is organized in the following way. In the next preliminary section, we introduce CAT(0) metric spaces, CAT(0) box and rectangular complexes, ramified rectilinear polygons, and squaregraphs. We also introduce the two-point shortest path query problem. In Section 3, we show that the shortest path $\gamma (x,y)$ between two points $x,y$ of a CAT(0) box complex $\mathcal K$ is contained in the subcomplex induced by the graph interval $I(p,q)$ between two vertices $p,q$ belonging to the cells containing $x$ and $y,$ respectively. Moreover, we show that this subcomplex  ${\mathcal K}(I(p,q))$ can be unfolded in the $k$-dimensional  space ${\mathbb R}^k$ (where $k$ is the dimension of a largest cell of ${\mathcal K}(I(p,q))$) in a such a way that the shortest path between any two points is the same in ${\mathcal K}(I(p,q))$ and in the unfolding of ${\mathcal K}(I(p,q)).$ In Section 4, we present the detailed description of the algorithm for answering two-point shortest path queries in  CAT(0) rectangular complexes and of the data structure $\mathcal D$ used in this algorithm. First we show how to compute the unfolding of ${\mathcal K}(I(p,q))$  in ${\mathbb R}^2$ efficiently. Then we describe the data structure $\mathcal D$ and show how to use it to compute the boundary paths of ${\mathcal K}(I(p,q)).$  $\mathcal D$ is different for general CAT(0) rectangular complexes, for ramified rectilinear polygons, and for squaregraphs. We conclude with a formal description of the algorithm and the analysis of its complexity.

\section{Preliminaries}

\subsection{CAT(0) metric spaces}
Let $(X,d)$ be a metric space. A \emph{geodesic}
joining two points $x$ and $y$ from $X$ is the image of a
(continuous) map $\gamma$ from a line segment $[0,l]\subset \mathbb{R}$
to $X$ such that $\gamma(0)=x, \gamma(l)=y$ and
$d(\gamma(t),\gamma(t'))=|t-t'|$ for all $t,t'\in [0,l].$ The space
$(X,d)$ is said to be \emph{geodesic} if every pair of points $x,y\in
X$  is joined by a geodesic \cite{BrHa}.  A \emph{geodesic triangle} $\Delta (x_1,x_2,x_3)$ in a geodesic
metric space $(X,d)$ consists of three distinct points in $X$ (the
vertices of $\Delta$) and a geodesic  between each pair of vertices
(the sides of $\Delta$). A \emph{comparison triangle} for $\Delta
(x_1,x_2,x_3)$ is a triangle $\Delta (x'_1,x'_2,x'_3)$ in the
Euclidean plane ${\mathbb E}^2$ such that $d_{{\mathbb
E}^2}(x'_i,x'_j)=d(x_i,x_j)$ for $i,j\in \{ 1,2,3\}.$ A geodesic
metric space $(X,d)$ is defined to be a \emph{$CAT(0)$ space} \cite{Gr}
if all geodesic triangles $\Delta (x_1,x_2,x_3)$ of $X$ satisfy the
comparison axiom of Cartan--Alexandrov--Toponogov:

\medskip\noindent
\emph{If $y$ is a point on the side of $\Delta(x_1,x_2,x_3)$ with
vertices $x_1$ and $x_2$ and $y'$ is the unique point on the line
segment $[x'_1,x'_2]$ of the comparison triangle
$\Delta(x'_1,x'_2,x'_3)$ such that $d_{{\mathbb E}^2}(x'_i,y')=
d(x_i,y)$ for $i=1,2,$ then $d(x_3,y)\le d_{{\mathbb
E}^2}(x'_3,y').$}

\medskip\noindent
This simple axiom turns out to be very powerful, because CAT(0)
spaces can be characterized  in several natural ways (for
a full account of this theory consult the book \cite{BrHa}). In particular,
a geodesic metric
space $(X,d)$ is CAT(0) if and only if any two points of this space
can be joined by a unique geodesic. CAT(0) is also equivalent to
convexity of the function $f:[0,1]\rightarrow X$ given by
$f(t)=d(\alpha (t),\beta (t)),$ for any geodesics $\alpha$ and
$\beta$ (which is further equivalent to convexity of the
neighborhoods of convex sets). This implies that CAT(0) spaces are
contractible.

\subsection{CAT(0) rectangular and box complexes}

A \emph{rectangular complex} $\mathcal K$ is a
2-dimensional cell complex $\mathcal K$ whose 2-cells are isometric to axis-parallel rectangles of
the $l_1$-plane. If all 1-cells of
$\mathcal K$ have equal length, then we call $\mathcal K$ a
\emph{square complex}; in this case we may assume without loss of generality that
the squares of the complex are all unit squares. Square complexes are the 2-dimensional
instances of \emph{cubical complexes,} viz. the cell complexes (where cells have
finite dimension)  in
which every cell of dimension $k$ is isometric to the unit cube of
${\mathbb R}^k.$ Analogously, the complexes in which all cells are axis-parallel boxes are high-dimensional generalizations
of rectangular complexes (we will call them {\it box complexes}).
 All complexes  occurring in our paper are finite, i.e., they have only finitely many cells.

The 0-dimensional faces of a rectangular or box complex $\mathcal K$ are called its
{\it vertices}, forming the vertex set $V({\mathcal K})$ of $\mathcal K$. The 1-dimensional faces of $\mathcal K$ are called the
{\it edges} of $\mathcal K$, and denoted by  $E({\mathcal K}).$ The {\it underlying graph} of $\mathcal K$ is the graph
$G({\mathcal K})= (V({\mathcal K}),E({\mathcal K})).$  Conversely, from any graph $G$
 one can derive a cube (or a box complex)  by replacing all subgraphs
of $G$ isomorphic to cubes of any dimensions
by solid cubes (or axis-parallel boxes). We denote any complex obtained in this way by $||G||$ and
call it  the {\it geometric
realization} of $G.$  A cell complex $\mathcal K$ is called \emph{simply connected} if it
is connected and every continuous mapping of the 1-dimensional
sphere ${\mathbb S}^1$ into $\mathcal K$ can be extended to a
continuous mapping of the disk ${\mathbb D}^2$ with boundary
${\mathbb S}^1$ into $\mathcal K$. The \emph{link} of a vertex $x$ in
$\mathcal K$ is the graph Link$(x)$ whose vertices are the 1-cells
containing $x$ and where two 1-cells are adjacent if and only if
they are  contained in  a common 2-cell (see \cite{BrHa} for the
notion of link in general polyhedral complexes). Given a subset $S$ of vertices of
$\mathcal K$, we will denote  by ${\mathcal K}(S)$ the subcomplex of ${\mathcal K}$
induced by $S$ and by $G(S)$ (or $G({\mathcal K}(S)$) the underlying graph of ${\mathcal K}(S).$

Computationally, a rectangular complex $\mathcal K$ is defined in the following way.
Each rectangular face $R$ of $\mathcal K$ is given by the circular list of four vertices and edges incident to $R.$ For each vertex
$v$ of  $\mathcal K$, the neighborhood of $v$ is given as the link graph  Link$(v);$ for each edge of Link$(v)$ there is a pointer
to the unique rectangular face containing $v$ and the edges of $\mathcal K$  incident to $v$ which define this edge of Link$(v).$
Finally, each point $x$ of $\mathcal K$ is given by its (local) coordinates in a rectangular cell $R(x)$ of $\mathcal K$ containing
$x$ (notice that $R(x)$ is unique if $x$ belongs to the interior of $R(x),$ otherwise $x$ may belong to several rectangular faces).

A rectangular or  box complex ${\mathcal K}$  can be endowed with several
intrinsic metrics \cite{BrHa} transforming ${\mathcal K}$ into a
complete geodesic space. Suppose that inside every cell of
${\mathcal K}$ the distance is measured according to an $l_1$- or $l_2$-metric. Then the
\emph{intrinsic} $l_1$- or $l_2$-\emph{metric}
of $\mathcal K$ is defined by assuming
that the distance between two points $x,y\in {\mathcal K}$ equals the infimum
of the lengths of the paths joining them. Here a \emph{path} in $\mathcal K$
from $x$ to $y$ is a sequence $P$ of points
$x=x_0,x_1\ldots x_{m-1}, x_m=y$ such that for each $i=0,\ldots,
m-1$ there exists a cell $R_i$ of $\mathcal K$ containing $x_i$ and $x_{i+1};$ the
\emph{length} of $P$ is $l(P)=\sum_{i=0}^{m-1} d(x_i,x_{i+1}),$ where
$d(x_i,x_{i+1})$ is computed inside $R_i$ according to the
respective metric. We denote the resulting $l_1$- and $l_2$-metrics on $\mathcal K$ by $d_1$ and $d_2,$ respectively.

The \emph{interval} between two points $x,y$ of a metric space $(X,d)$ is the set $I(x,y)=\{ z\in
X: d(x,y)=d(x,z)+d(z,y)\}$; for example, in Euclidean spaces, the interval $I(x,y)$ is the closed line segment having $x$ and $y$ as its endpoints.
A subspace $Y$ of a metric space $(X,d)$ is \emph{gated} if for every
point $x\in X$ there exists a (unique) point $x'\in Y,$ the \emph{gate} of $x$ in $Y$,  such that $d(x,y)=d(x,x')+d(x',y)$ for all
$y\in Y.$   Gated subspaces are necessarily convex, where a subspace $Y$ of $(X,d)$ is called \emph{convex} if $I(x,y)\subseteq Y$ for any $x,y\in Y$.
A \emph{half-space} $H$ of $X$ is a convex
subspace with a convex complement. For three points $x,y,z$ of a metric space $(X,d),$ let  $m(x,y,z)=I(x,y)\cap I(y,z)\cap I(z,x).$
If $m(x,y,z)$ is a singleton for all $x,y,z\in X$, then the space
$X$ is called \emph{median} \cite{BaCh_survey,VdV} and we usually refer to
$m(x,y,z)$ as to the \emph{median} of $x,y,z$ (here we do not
distinguish between the singleton and the corresponding point). A graph $G$ is a \emph{median graph} if
$(V,d_G)$ is a median space, where $d_G$ is the standard graph-metric of $G$.
Discrete median spaces, in general, can be regarded as median
networks: a \emph{median network} is a median graph with weighted
edges such that opposite edges in any 4-cycle have the same length \cite{Ba_condorcet}. Median
graphs not containing any induced cube (or cube network,
respectively) are called \emph{cube-free.} A {\it median complex} is a cube or a box complex of a median graph.
We will say that a subset $S$ of points of a CAT(0) rectangular complex $\mathcal K$ is $d_i$-{\it convex} for $i=1,2,$ if $S$ is a convex
subset of the metric space $({\mathcal K},d_i)$ (convex subsets of the
underlying graph $G({\mathcal K})$ will be called {\it graph-convex}).

Now we recall the combinatorial characterization of CAT(0) cubical and box complexes given by  Gromov.

\begin{theorem} \label{Gr} \cite{Gr} A cubical (or a box) polyhedral complex
${\mathcal K}$ with the intrinsic $l_2$-metric $d_2$ is CAT(0) if and
only if ${\mathcal K}$ is simply connected and  satisfies the
following condition: whenever three $(k + 2)$-cubes of ${\mathcal
K}$ share a common $k$-cube and pairwise share common $(k
+1)$-cubes, they are contained in a $(k+3)$--cube of ${\mathcal
K}.$
\end{theorem}

In some recent papers, CAT(0) cubical polyhedral complexes were
called cubings \cite{Sa}. With some abuse of language, we will call {\it cubings}
all CAT(0) box complexes.   The following relationship holds between
cubings and median polyhedral complexes (this result was used in several recent papers in
geometric group theory \cite{ChDrHa,Ni}).

\begin{theorem} \label{cubings_median} \cite{Ch_CAT,Ro} Median complexes  and cubings
(both equipped with the $l_2$-metric) constitute the same
objects.
\end{theorem}

In this paper we will mainly investigate the CAT(0) rectangular complexes (i.e., 2-dimensional cubings), which can be characterized in the following way:

\begin{theorem} \cite{BaChEp_ramified} For a rectangular complex $\mathcal K$ the following conditions  are
equivalent:
\begin{itemize}
\item[(i)] the underlying graph $G({\mathcal K})$ of ${\mathcal K}$ is a cube-free median graph;
\item[(ii)] the metric space $(\mathcal K,d)$ is median;
\item[(iii)] $\mathcal K$ equipped with the intrinsic $l_2$-metric $d_2$ is CAT(0);
\item[(iv)] $\mathcal K$ is simply connected and for every vertex
$x\in V({\mathcal K}),$ the graph Link$(x)$ is triangle-free.
\end{itemize}
\end{theorem}

\begin{figure}\centering
\begin{tabular}{cc}
   \includegraphics[width=0.35\textwidth]{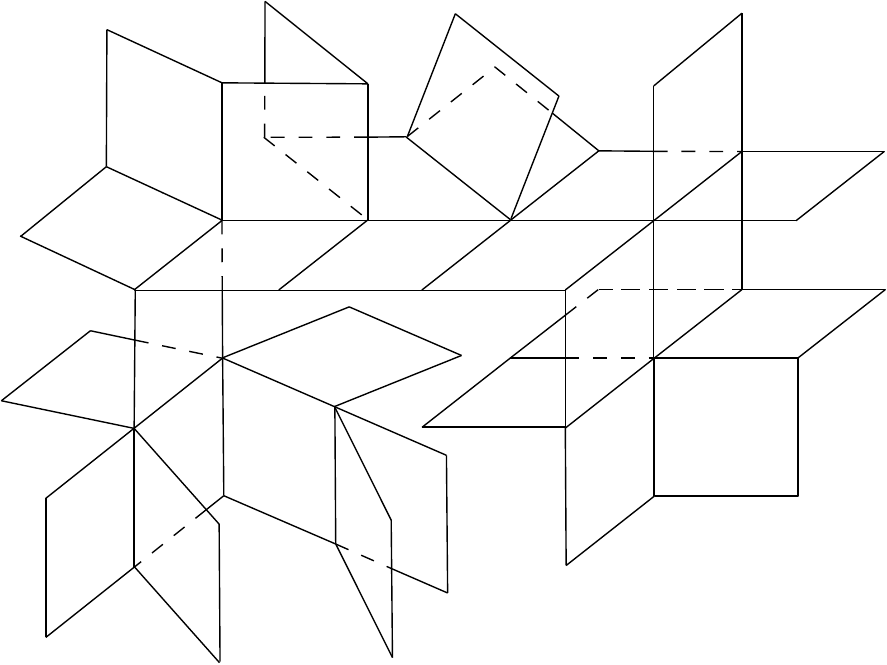}
   &
   \includegraphics[width=0.35\textwidth]{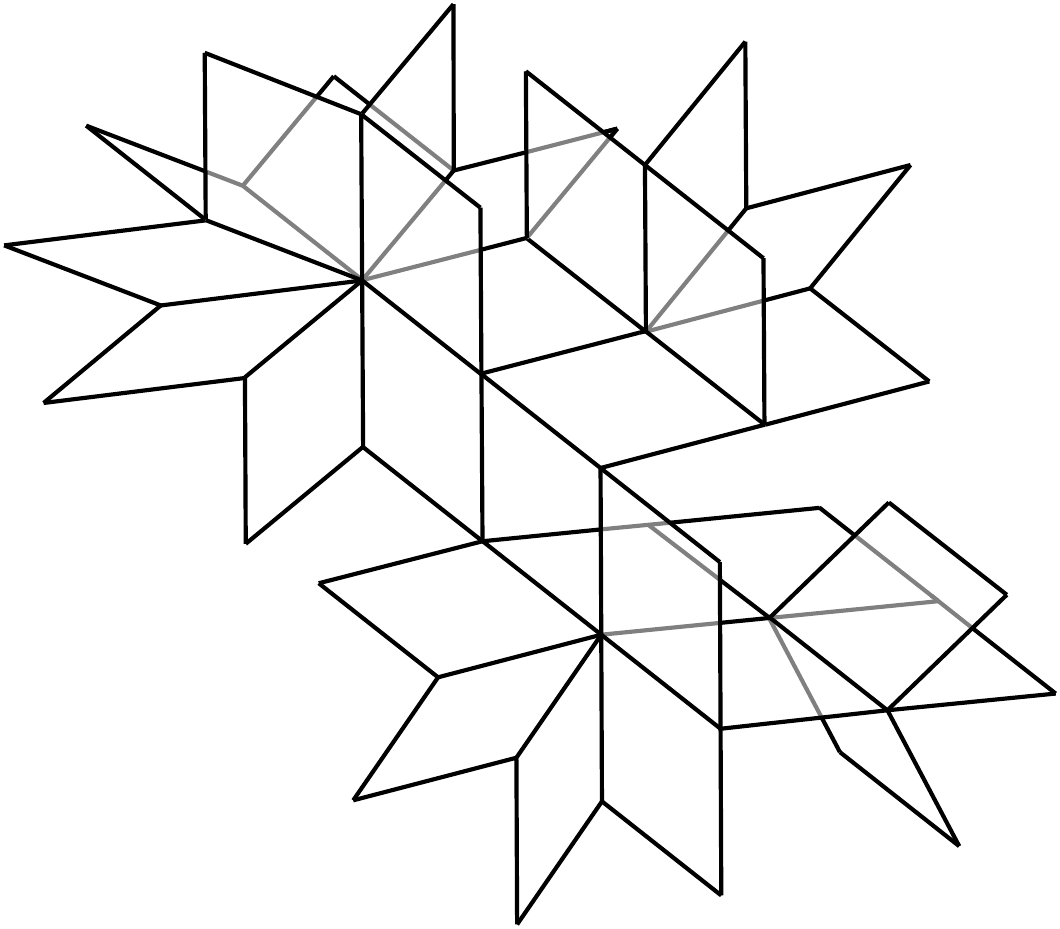}
   \\
   (a)
   &
   (b)
\end{tabular}
\includegraphics[width=0.35\textwidth]{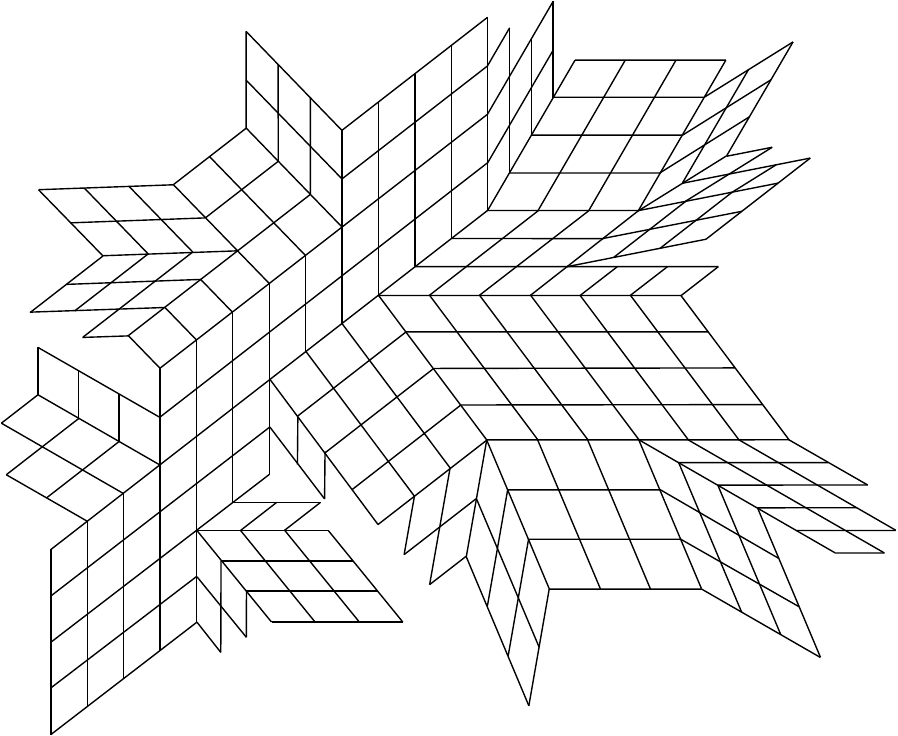}
\\
 (c)
 \caption{(a) a CAT(0) rectangular complex, (b) a ramified rectilinear polygon, (c) a squaregraph.} \label{des_ex}
\end{figure}

Typical examples of CAT(0) rectangular complexes (which are illustrated in  Fig. \ref{des_ex}) are  the
{\it squaregraphs} (i.e., the rectangular complexes obtained from the plane graphs in which all inner faces are
4-cycles and all inner vertices have degrees $\ge 4$ \cite{BaChEp_square,ChDrVa_soda}) and the {\it ramified rectilinear polygons}
(i.e., rectangular complexes   endowed with the intrinsic
$l_1$-metric which embeds isometrically into the product of two
finite trees \cite{BaChEp_ramified}).  As is established in \cite{BaChEp_ramified}, ramified rectilinear polygons are exactly
the simply connected rectangular complexes $\mathcal K$ in which the graph Link$(x)$ is bipartite  for each vertex $x$ of ${\mathcal K}.$ 

In median graphs, the halfspaces (the convex sets with convex complements) have a special structure and plays an important role.
It is well known \cite{BaVdV,Mu1,Mu} that median
graphs isometrically embed into hypercubes.  The isometric embedding of a median graph $G$ into a (smallest)
hypercube coincides with the so-called canonical embedding, which is
determined by the Djokovi\'c-Winkler relation $\Theta$ on the edge
set of $G:$ two edges $uv$ and $wx$ are $\Theta$-related exactly
when $d_G(u,w)+d_G(v,x)\ne d_G(u,x)+d_G(v,w);$ see \cite{EppFaOv,ImKl}. For a
median graph this relation is transitive and hence an equivalence
relation. It is the transitive closure of the ``opposite" relation
of edges on 4-cycles (i.e., 2-dimensional faces of $\mathcal K$):
in fact, any two $\Theta$-related edges can be
connected by a ladder (viz., the Cartesian product of a path with
$K_2$), and all edges $\Theta$-related to some edge
$uv$ constitute a cutset $\Theta(uv)$ of the median graph, which
determines one factor of the canonical hypercube \cite{Mu1}. The
cutset $\Theta(uv)$ defines two complementary halfspaces
(convex sets with convex complements) $W(u,v),W(v,u)$
of $G$ \cite{Mu,VdV}, where $W(u,v)=\{ x\in X:
d(u,x)<d(v,x)\}$ and $W(v,u)=V-W(u,v).$  Conversely, for any pair of complementary halfspaces $H_1,H_2$
of a median graph $G$ there exists an edge $xy$ such
that $W(x,y)=H_1$ and $W(y,x)=H_2$ is the given pair of halfspaces (in fact, all edges belonging to the same equivalence
$\Theta$-class as $xy$ define the same pair of complementary halfspaces).

\begin{figure}\centering
\begin{tabular}{cc}
   \includegraphics[width=0.4\textwidth]{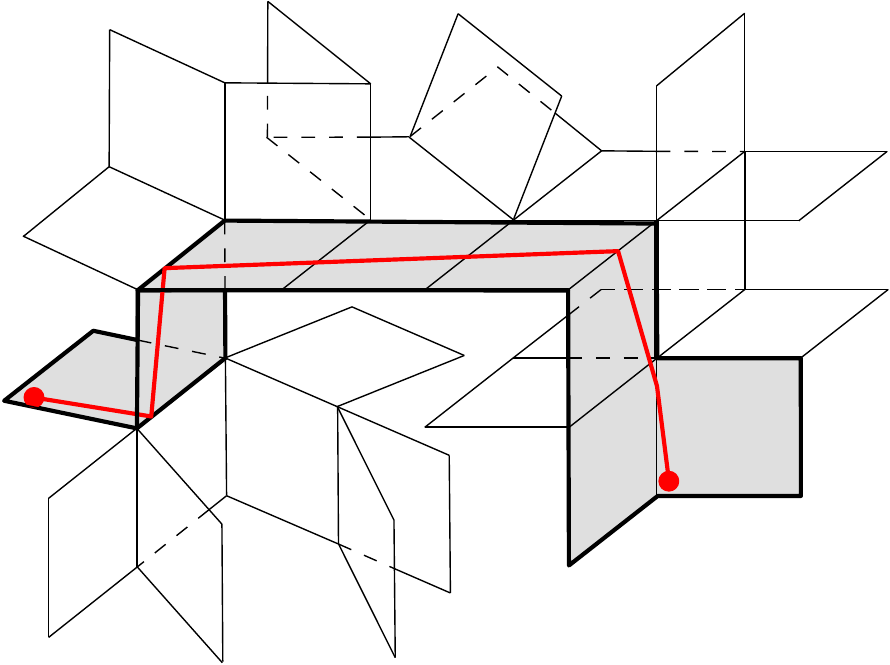}
   &
   \includegraphics[width=0.55\textwidth]{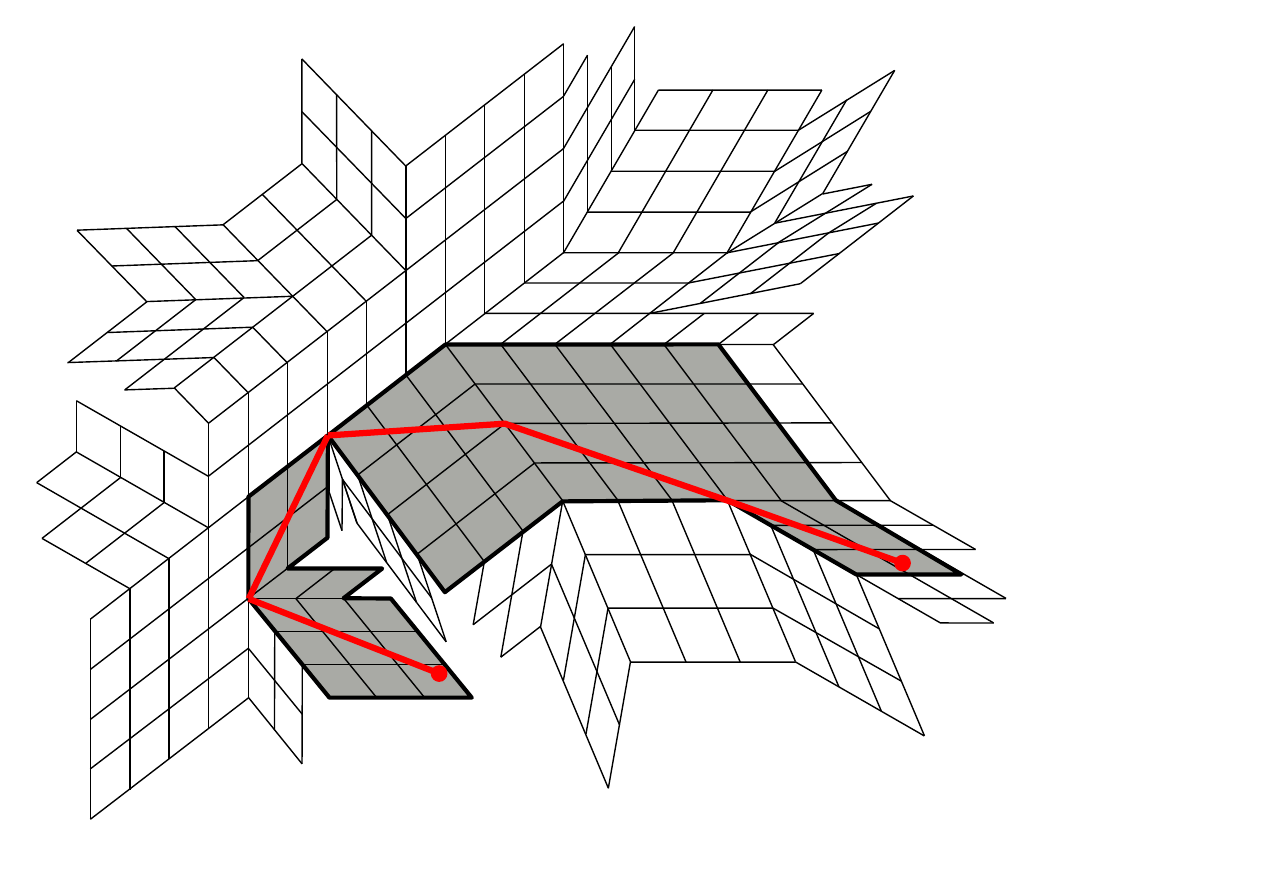}
   \\
   (a)
   &
   (b)
\end{tabular}
\caption{Two-point shortest path queries.} \label{des_interval}
\end{figure}

In this paper, we consider the following shortest path problem in CAT(0) rectangular complexes ${\mathcal K}$ endowed with the intrinsic $l_2$-metric $d_2$ (we illustrate this formulation in Fig. \ref{des_interval}):

\medskip\noindent
{\bf Two-point queries:} Given two points $x,y$ of ${\mathcal K},$ compute the unique shortest path $\gamma(x,y)$ between $x$ and $y$ in $\mathcal K$.

%


\section{Geodesics and graph-intervals}

In this section, we show that, given two arbitrary points $x,y$ of a CAT(0) box complex ${\mathcal K},$ the $d_2$-shortest path $\gamma (x,y)$ is always contained in the subcomplex induced by the graph interval $I(p,q)$ between two vertices $p,q$ belonging to the cells containing $x$ and $y,$ respectively.  Moreover, we show that this subcomplex  ${\mathcal K}(I(p,q))$ can be unfolded in the $k$-dimensional  space ${\mathbb R}^k$ (where $k$ is the dimension of a largest cell of ${\mathcal K}(I(p,q))$) in a such a way that the $d_2$-shortest path between any two points is the same in ${\mathcal K}(I(p,q))$ and in the unfolding of ${\mathcal K}(I(p,q)).$


\begin{proposition} \label{interval_vertices} If $p$ and $q$ are two vertices of a CAT(0) box complex  $\mathcal K$, then ${\mathcal K}(I(p,q))$ is $d_2$-convex and therefore $\gamma(x,y)\subset {\mathcal K}(I(p,q))$ for any two points $x,y\in {\mathcal K}(I(p,q)).$
\end{proposition}

\begin{proof}  According to Theorem \ref{cubings_median}, CAT(0) box complexes are exactly the box complexes  having median graphs as underlying graphs. Let $G=G(\mathcal K)$ be the underlying graph of $\mathcal K$.  Since $G$ is a median graph, the interval $I(p,q)$ is a convex subset (and therefore a gated subset) of $G$ \cite{VdV}. Additionally, in median graphs each convex set $S$ can be written as an intersection of halfspaces \cite{VdV} (we will present a simple proof of this fact below - Lemma \ref{convex_set}). Therefore, it suffices to show that for each  halfspace $H$ of the graph $G,$ the subcomplex $\mathcal K (H)$ is $d_2$-convex. Indeed, this will show that ${\mathcal K}(S)$ can be represented as an intersection of $d_2$-convex sets of $\mathcal K$ and therefore that ${\mathcal K}(S)$ itself is $d_2$-convex.


\begin{lemma} \label{convex_set}
A convex set $S$ of a median graph $G$ is the intersection of the half-spaces containing $S.$
\end{lemma}
\begin{proof}
In order to prove the lemma, it suffices to show that for any vertex $v$ of $G$ not belonging to $S$ there exists a pair of complementary halfspaces $H_1,H_2$ of $G$ such that $v\in H_1$ and $S\subseteq H_2.$ Since $S$ is convex and $G$ is median, $S$ is gated \cite{VdV}. Let $u$ be the gate of $v$ in $S$ (i.e., $u\in I(v,x)$  for each vertex $x\in S$). Let $v'$ be a neighbor of  $v$ in the interval $I(v,u).$ Consider the complementary halfspaces $W(v,v')$ and $W(v',v)$ of $G$ defined by the edge $vv'.$ Then obviously $v\in W(v,v').$ On the other hand, since $v'\in I(v,u)\subseteq I(v,x)$ for any vertex $x\in S,$ by the definition of $W(v',v)$ we conclude that $x\in W(v',v),$ yielding $S\subseteq W(v',v)$ and concluding the proof.
\end{proof}

\begin{figure}\centering
\begin{tabular}{cc}
   \includegraphics[width=0.4\textwidth]{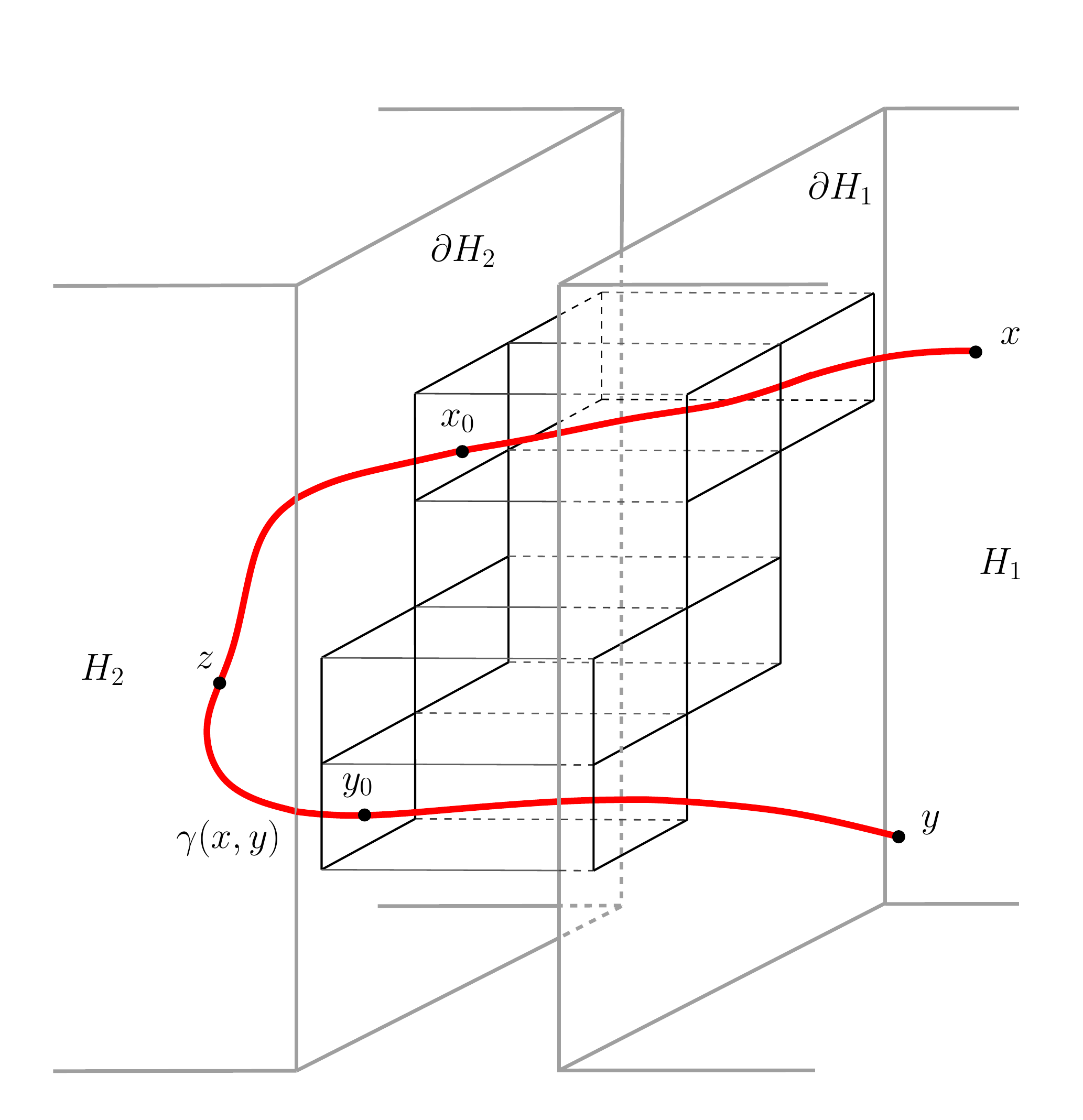}
   &
   \includegraphics[width=0.4\textwidth]{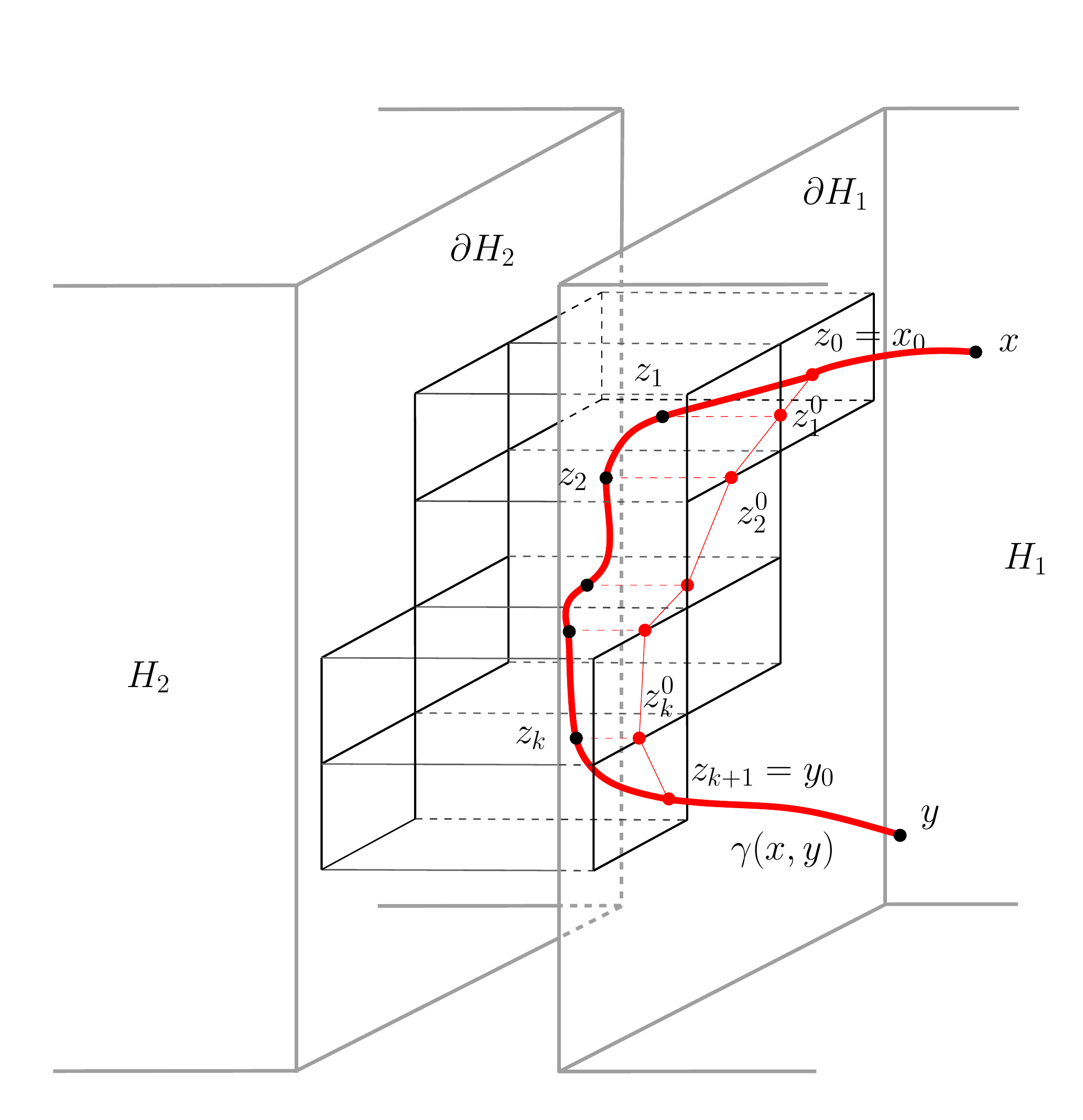}
   \\
   (a)
   &
   (b)
\end{tabular}
\caption{To the proof of Proposition \ref{interval_vertices}.} \label{des_interval_vertices}
\end{figure}

Finally, we will prove now that for any pair of complementary halfspaces $H_1,H_2$ of
$G$ the subcomplexes ${\mathcal K}(H_1)$ and ${\mathcal K}(H_2)$ are $d_2$-convex. Suppose without loss of generality that $H_1$ and $H_2$ are defined by
the edges of the equivalence class $\Theta_i$ of $\Theta$. The {\it boundary} $\partial H_1$ of $H_1$
consists of all ends of edges of $\Theta_i$ belonging to $H_1$  (the boundary $\partial H_2$ of $H_2$ is defined in a similar way). It was shown by Mulder \cite{Mu1,Mu}
that in median graphs $G$  the boundaries $\partial H_1$ and $\partial H_2$ of complementary halfspaces $H_1,H_2$ induce isomorphic convex, and therefore median,
subgraphs of $G$.  Hence ${\mathcal K} (\partial H_1)$ and
${\mathcal K} (\partial H_2)$ are isomorphic CAT(0) subcomplexes of ${\mathcal K},$ which we call {\it hyperplanes}. Note also that $H_1\cup {\partial H_2}$
and $H_2\cup {\partial H_1}$ induce convex, and therefore median, subgraphs of $G.$  All edges $xy$ of $\Theta_i$ have
the same length $l_i$ in $\mathcal K$. The CAT(0) subcomplex ${\mathcal K}(\partial H_1\cup \partial H_2)$ of $\mathcal K$ is isomorphic to the CAT(0) box complex
${\mathcal K}_i=\partial H_1\times[0,l_i].$ We will show that ${\mathcal K}(H_1)$ and ${\mathcal K}(H_2)$ are $d_2$-convex by induction on the number of vertices of
$\mathcal K$. Suppose by way of contradiction that ${\mathcal K}(H_1)$ is not $d_2$-convex. Then there exist two points $x,y\in {\mathcal K}(H_1)$ such that the geodesic
$\gamma(x,y)$ has a point which does not belong to ${\mathcal K}(H_1).$ First suppose that $\gamma(x,y)$ contains a point $z$ which belongs to
${\mathcal K}(H_2\setminus \partial H_2);$ see Fig. \ref{des_interval_vertices}(a) for an illustration.  Then we can find two points $x_0\in \gamma(x,z)\cap {\mathcal K}({\partial H_2})$ and $y_0\in \gamma(y,z)\cap {\mathcal K}({\partial H_2}).$
Then $z\in \gamma(x_0,y_0),$ showing that ${\mathcal K}({\partial H_2})$ is not $d_2$-convex in $\mathcal K$ and in the CAT(0) subcomplex  ${\mathcal K}(H_2).$ Since $\partial H_2$ is
a convex subgraph of $G$ and therefore of the median subgraph $G(H_2)$ of $G$ induced by $H_2,$ we conclude that in the CAT(0) complex  ${\mathcal K}(H_2)$ not every convex subgraph induces
a $d_2$-convex subcomplex, contrary to the induction hypothesis.

Therefore, we can assume that the geodesic $\gamma(x,y)$ is entirely contained in the subcomplex of $\mathcal K$
induced by $H_1\cup \partial H_2;$ this case is illustrated in Fig. \ref{des_interval_vertices}(b). Since $\gamma(x,y)$ is a closed set of $\mathcal K$, we can find two (necessarily different) points $x_0,y_0\in \gamma(x,y)\cap {\mathcal K}(\partial H_1)$ such that all points of the geodesic $\gamma(x_0,y_0)$ except $x_0$ and $y_0$ all belong to the strip  ${\mathcal K_i}$ minus the hyperplane $\mathcal K (\partial H_1).$ Let $C_0,C_1,\ldots,C_k$ be the sequence of maximal by inclusion cells of  ${\mathcal K_i}$ intersected by $\gamma(x_0,y_0)$ labeled in the order in which they intersect $\gamma(x_0,y_0)$ (so that $x_0\in C_0$ and
$y_0\in C_k$). Let $z_i\in \gamma(x_0,y_0)\cap C_{i-1}\cap C_{i},$ $i=1,\ldots,k,$ and set $z_0=x_0,z_{k+1}=y_0.$ Then each of the geodesics $\gamma(z_{i-1},z_i)$ belongs to the cell $C_{i-1},$ respectively.
The intersection $C_i^0$ of each cell $C_i$ with the hyperplane ${\mathcal K}({\partial H_1})$ is a cell of ${\mathcal K}({\partial H_1})$ and  also a facet of $C_i.$
The orthogonal projection $\pi_i$ of each box $C_i$ $(i=0,\ldots,k)$ on its facet $C^0_i$ is a non-expansive map (with respect to the $d_2$-metric). Notice that $\pi_0(z_0)=z_0=y_0$ and $\pi_k(z_{k+1})=z_{k+1}=y_0.$ On the
other hand,  since $z_i\in C_{i-1}\cap C_{i}$ and the cells $C_{i-1}$ and $C_i$ are axis-parallel boxes, we conclude that the two projections  $\pi_{i-1}(z_i)$ and $\pi_i(z_i)$ are
one and the same point $z^0_i$ of the cell $C^0_{i-1}\cap C^0_i\subset C_{i-1}\cap C_{i}.$   Consider the path $\gamma^0(x_0,y_0)$ between $x_0$ and $y_0$ in the
hyperplane ${\mathcal K}({\partial H_1})$ obtained by concatenating  the geodesics $\gamma(z^0_0=x_0,z^0_1),\gamma(z^0_1,z^0_2),\ldots, \gamma(z^0_k,y_0=z^0_{k+1}).$ Since for each $i=0,1,\ldots,k,$ the map $\pi_i$  is non-expansive on $C_i$,  the length of the geodesic $\gamma(z^0_{i},z^0_{i+1})$ is less or equal to the length of the geodesic $\gamma(z_{i},z_{i+1}).$ Therefore  $\gamma^0(x_0,y_0)$ is a geodesic between $x_0$ and $y_0$ which is completely contained in the hyperplane ${\mathcal K}({\partial H_1}).$ Since $\gamma(x_0,y_0)\cap {\mathcal K}({\partial H_1})=\{ x_0,y_0\},$ we conclude that $x_0$ and $y_0$ are connected in $\mathcal K$  by two different geodesics $\gamma(x_0,y_0)$ and $\gamma^0(x_0,y_0)$, contrary to the assumption that $\mathcal K$ is CAT(0). This contradiction establishes that the halfspaces of $G$ induce indeed $d_2$-convex subcomplexes of $\mathcal K$, establishing in particular that ${\mathcal K}(I(p,q))$ is $d_2$-convex for any two vertices $p,q$ of $\mathcal K$. 
$\Box$
\end{proof}

\begin{proposition} \label{interval_point} If $x$ and $y$ are two arbitrary points of a CAT(0) box complex $\mathcal K,$ and $R(x)$ and $R(y)$ are two minimal by inclusion cells of $\mathcal K$ containing $x$ and $y,$ respectively, then $\gamma(x,y)\subset {\mathcal K}(I(p,q)),$ where $p$ and $q$ are mutually furthest (in the graph $G({\mathcal K})$) vertices of $R(x)$ and $R(y).$
\end{proposition}

\begin{proof}  $R(x)$ and $R(y)$ are the unique cells of least dimension such that $x$ belongs to the relative interior of $R(x)$ and $y$ belongs to the relative interior of $R(y).$ The sets of vertices of $R(x)$ and $R(y)$ are convex, and therefore gated, subsets of $G.$ Let $p\in R(x)$ and $q\in R(y)$ be two mutually furthest vertices of $R(x)$ and $R(y),$ i.e. $d(p,q)=\max\{ d(p',q'): p'\in V(R(x)), q'\in V(R(y))\},$ where all distances $d(p',q')$ are computed according to the graph-distance in $G({\mathcal K}).$ Since $G$ is bipartite, the choice of the pair $p,q$ implies that all neighbors of $p$ in the graphic cube $G(R(x))$ must be one step closer to $q$ than  $p,$ i.e., all these vertices (we denote this set by $A$) belong to the interval $I(p,q).$ Analogously, all neighbors of $q$ in $G(R(y))$ (we denote this set by $B$)  also belong to the interval $I(p,q).$ Since $I(p,q)$ is convex and the convex hull of $A$ in $G$ contains the whole graphic cube $G(R(x))$ while the convex hull of $B$ contains $G(R(y)),$ we conclude that both $R(x)$ and $R(y)$ belong to the subcomplex ${\mathcal K}(I(p,q)).$ Since by Proposition \ref{interval_vertices}, ${\mathcal K}(I(p,q))$ is $d_2$-convex  and $x\in R(x), y\in R(y),$ we conclude that $\gamma(x,y)\subset {\mathcal K}(I(p,q)).$ $\Box$
\end{proof}

The next result shows that the intervals $I(p,q)$ in the CAT(0) box complexes can be unfolded in  Euclidean spaces of dimension equal to the topological dimension (i.e., the least dimension of a cell) of ${\mathcal K}(I(p,q)).$   Recall that a function $f:
X\rightarrow X'$ between two metric spaces $(X,d)$ and $(X',d')$ is
an  \emph{isometric embedding} of $X$ into $X'$ if
$d'(f(x),f(y))=d(x,y)$ for any $x,y\in X.$ In this case $Y:=f(X)$ is
called an \emph{(isometric) subspace} of $X'.$ If a mapping $f:X\mapsto X'$ between two geodesic metric spaces $(X,d)$ and $(X',d')$ is such that $f(X)$ is geodesic and compact, then $f(X)$ is called an {\it unfolding} of $X$ in $X'$ if $f$ is an isometric embedding of $(X,d)$ in $(f(X),d^*)$, where $d^*$ is the intrinsic metric on $f(X)$ induced by $d'.$

\begin{proposition} \label{interval_zk}  If $p$ and $q$ are two vertices of a CAT(0) box complex $\mathcal K$ and  $k$ is the largest dimension of a cell of ${\mathcal K}(I(p,q))$, then there exists an embedding $f^*$ of ${\mathcal K}(I(p,q))$  in the $k$-dimensional Euclidean space ${\mathbb R}^k$ such that $f^*({\mathcal K}(I(p,q)))$ is an unfolding of ${\mathcal K}(I(p,q)).$
\end{proposition}

\begin{proof} First, we show that the subgraph $G(I(p,q))$ of $G=G({\mathcal K})$ induced by the interval $I(p,q)$ can be isometrically embedded in the $k$-dimensional cubical grid ${\mathbb Z}^k=\Pi _{i=1}^k P_i,$ where each $P_i$ is the infinite path having ${\mathbb Z}$ as the set of vertices. Indeed,
intervals $I(p,q)$ of median graphs and median semilattices can be viewed as  distributive lattices by setting  $x\wedge y=m(p,x,y)$ and $x\vee y=m(q,x,y)$ for any  $x,y\in I(p,q),$ where $m$ is the median operator of $G$ \cite{BaHe,BiKi}. Using the encoding of distributive lattices via closed subsets of a poset due to Birkhoff \cite{Bi}, the famous Dilworth's theorem (the size of a largest antichain
of a poset equals to the least size of a  decomposition of the poset into chains) \cite{Dil} implies that any distributive lattice $L$ of breadth $k$ can be embedded as a sublattice of a product of $k$ chains, see \cite{La} or \cite{ChSu}  for this interpretation of Dilworth's result (the breadth of a distributive lattice $L$ is equal to the largest out- or in-degree of a vertex in the covering graph of $L$). Larson \cite{La} showed that the resulting embedding can be chosen
to preserve the covering relation, i.e. to be a graph embedding. Recently, using the same tools,  Cheng and Suzuki \cite{ChSu} noticed that the embedding can be selected to be an isometric
embedding of the covering graph of a distributive lattice of breadth $k$ in the product of $k$ chains (note that Eppstein  \cite{Epp_lattice} showed how to decide in polynomial time if a
graph $G$ isometrically embeds into the product of $k$ chains).

Therefore, it remains to show that the largest out-degree or in-degree of $G(I(p,q))$ equals to the dimension of a largest cube of
$G(I(p,q)).$ For this, it suffices to show that if $v\in I(p,q),$ then any $m$ neighbors $y_1,y_2,\ldots,y_m\in I(v,q)\subseteq I(p,q)$ of $v$ define an $m$-cube $C_m\subseteq I(v,q).$ We
proceed by induction on $m.$  Denote by $C'$ the $(m-1)$-cube induced by the vertices $y_1,\ldots,y_{m-1}$ (which
exists because of the induction assumption). Let $z_i$ be the median of the triplet $y_i,y_m,q.$ Then $z_1,\ldots,z_{m-1}$ are all adjacent to $y_m$ and therefore are pairwise different
(because median graphs are $K_{2,3}$-free),  and all belong to the interval $I(y_m,q).$ Therefore, by induction hypothesis,  $z_1,\ldots,z_{m-1}$ induce an $(m-1)$-cube $C''\subseteq I(y_m,q).$
Then it can be easily shown that each vertex of $C'$ is adjacent to a unique vertex of $C''$ and that this adjacency relation induces  an isomorphism between the cubes $C'$ and $C''$.
Hence $C'\cup C''$ is an $m$-dimensional cube. Thus indeed  $G(I(p,q))$  can be isometrically embedded in the $k$-dimensional grid ${\mathbb Z}^k.$

Denote by $f$ such an embedding. To transform $f$ into an unfolding of ${\mathcal K}(I(p,q))$ in ${\mathbb R}^k,$ we simply transform the uniform cubical grid  ${\mathbb Z}^k$ into a non-uniform one: notice that all edges of
the same equivalence class $\Theta_i$ of $G(I(p,q))$ are mapped by $f$ to one and the same edge $e$ of some path $P_j.$  If the edges of $\Theta_i$ all have length $l_i,$ then we simply
assign length $l_i$ to the edge $e$ of $P_j.$ Denote the resulting paths by $P^*_1,\ldots,P^*_k$ and notice that after this scaling the previous embedding induce an embedding $f^*$ of the graph
$G(I(p,q))$ weighted by the length of edges in ${\mathcal K}$ into the grid $\Pi _{i=1}^k P^*_i.$ We can extend in a natural way $f^*$ to an embedding of ${\mathcal K}(I(u,v))$ into
${\mathbb R}^k=||\Pi _{i=1}^k P^*_i||:$ for a cell $R$ of  ${\mathcal K}(I(u,v)),$ $f^*(R)$ is the cell induced by the images under $f^*$  of the vertices of $R.$ Let $f^*({\mathcal K}(I(p,q)))$
denote the box complex consisting of the images of all cells of ${\mathcal K}(I(p,q)).$ Since each path between two points $x,y$ of  ${\mathcal K}(I(p,q))$ is mapped to a path of the same length of  $f^*({\mathcal K}(I(p,q)))$ between $f^*(x)$ and $f^*(y)$, we obtain the desired unfolding of  ${\mathcal K}(I(p,q))$  in the $k$-dimensional Euclidean space. $\Box$
\end{proof}

For efficient (but nonlinear) algorithms for isometric embeddings of median graphs into cubical grids of least dimension, see the recent paper by Cheng \cite{Cheng}.

\section{Two-point shortest path queries}

In this section, we present the detailed description of the algorithm for answering two-point shortest path queries in  CAT(0) rectangular complexes and of the data structure $\mathcal D$ used in this algorithm. First we show that  ${\mathcal K}(I(p,q))$ always can be  unfolded in the plane as a chain of monotone polygons, which we will denote by  $P(I(p,q)),$ and we show how to compute this unfolding efficiently.   Therefore, to compute  $\gamma(x,y),$ we triangulate each monotone polygon of $P(I(p,q))$  and compute in linear time the shortest path $\gamma^*(x,y)$ in $P(I(p,q))$ between the images of $x$ and $y$ (we denote them also by $x,y$) using the algorithm of
Lee and Preparata \cite{LeePre} and return as $\gamma(x,y)$ the preimage of $\gamma^*(x,y).$ As a  preprocessing step, we design a data structure $\mathcal D$  allowing for each query $x,y$ to efficiently retrieve  the boundary of an interval $I(p,q)$ such that $x,y\in {\mathcal K}(I(p,q))$ and $x,p$ and $y,q$ belongs to common rectangular cells, respectively (in time proportional to the distance $d(p,q)$ between $p$ and $q$ in $G({\mathcal K})$).

\subsection{The unfolding of ${\mathcal K}(I(p,q))$ in ${\mathbb R}^2$}

From Proposition \ref{interval_zk}, we know that for any two vertices $p$ and $q$ of a CAT(0) rectangular complex $\mathcal K$, the graph $G(I(p,q))$ is isometrically embeddable in ${\mathbb Z}^2$ and consequently the subcomplex ${\mathcal K}(I(p,q))$ can be unfolded in ${\mathbb R}^2.$ We will show how to compute such an unfolding efficiently. Denote by $P(I(p,q))$ the image of ${\mathcal K}(I(p,q))$ under an unfolding $f$ (which we know to exist).  Let $B_1,\ldots, B_m$ be the 2-connected components (alias blocks) of the graph $G(I(p,q)).$ From the definition of $I(p,q),$ it follows that  $B_1,\ldots, B_m$ define a chain of blocks, i.e., if $s_1\in B_1\cap B_2, \ldots, s_{m-1}\in B_{m-1}\cap B_m$ are the articulation vertices of $G(I(p,q))$ and $s_0=p, s_m=q,$  then $B_i=I(s_{i-1},s_i),$ $i=1,\ldots,m,$ and all vertices $s_1,\ldots, s_{m-1}$ belong to all shortest paths between $p$ and $q.$  In the squaregraph $G(I(p,q))$ each hyperplane is a convex  path and in any isometric embedding of $G(I(p,q))$ in ${\mathbb Z}^2,$ the image of this path is a horizontal or a vertical path. Therefore $P(I(p,q))=\cup_{i=1}^m P_i$ is horizontally and vertically convex (i.e., the intersection of $P(I(p,q))$ with each horizontal or vertical line is convex) and consists  of a chain of monotone polygons $P_1=f({\mathcal K}(B_1)),\ldots, P_m=f({\mathcal K} (B_m)).$ Up to translations, rotations, and symmetries, each block $B_i$ has a unique isometric embedding in ${\mathbb Z}^2,$ whence each ${\mathcal K}(B_i)$ has a unique unfolding in the plane. We will define the {\it boundary} $\partial G(I(p,q))$ of $G(I(p,q))$ (or of ${\mathcal K}(I(p,q))$) as the subgraph induced by all edges of $G(I(p,q))$ which are mapped to the boundary $\partial P(I(p,q))$ of $P(I(p,q))$ (since $G(I(p,q))$ is a squaregraph, this definition is equivalent to the definition of a boundary of a squaregraph given in \cite{BaChEp_square}). Given two arbitrary points $x,y\in {\mathcal K}(I(p,q)),$ it is a well-known property of simple polygons that all vertices  of the (Euclidean) shortest path in $P(I(p,q))$ between $f(x)$ and $f(y)$ are vertices of the boundary of $P(I(p,q))$(except $f(x)$ and $f(y)$). Therefore, instead of defining the image of the whole subcomplex ${\mathcal K}(I(p,q))$ under an unfolding map $f,$ it suffices to define only the
image under $f$  of its boundary $\partial G(I(p,q))$  (in this case, we will speak about the unfolding of  $\partial G(I(p,q))$).

\begin{figure}\centering
\includegraphics[width=0.9\textwidth]{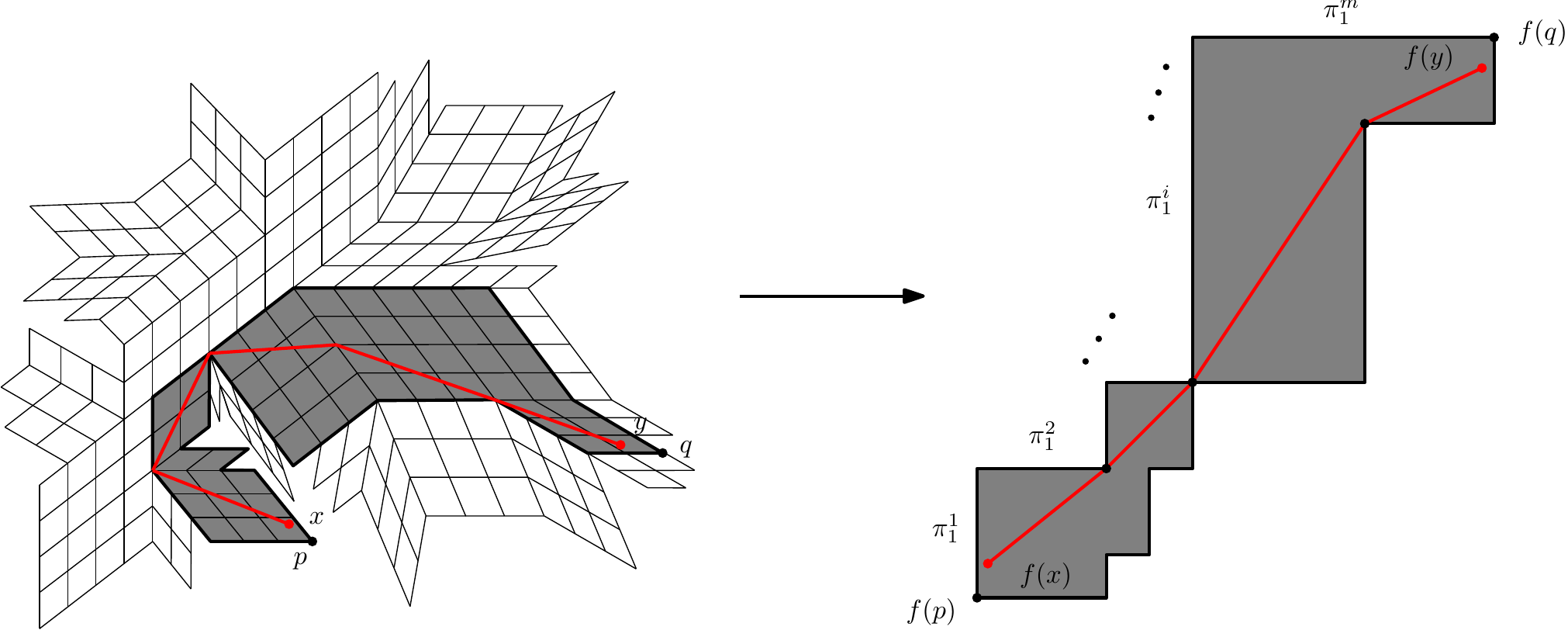}
\caption{An unfolding of ${\mathcal K}(I(p,q))$ into blocks.} \label{2-points query}
\end{figure}

\begin{proposition} \label{interval_z2}  If $p$ and $q$ are two vertices of a CAT(0) rectangular complex $\mathcal K,$  then ${\mathcal K}(I(p,q))$ can be unfolded in the Euclidean plane as a chain of monotone polygons and this embedding can be constructed in $O(|I(p,q)|)$ time if $I(p,q)$ is given.  Moreover, if the boundary $\partial G(I(p,q))$ is given together with the degrees $deg_0(z)$ in $G(I(p,q))$ of all of its vertices $z$, then an unfolding of  $\partial G(I(p,q))$ can be constructed in  $O(d(p,q))$ time,   where $d(p,q)$ is the distance between the vertices $p$ and $q$ in the graph $G({\mathcal K}).$
\end{proposition}

\begin{proof}  Suppose that either $I(p,q)$
or its boundary $\partial  G(I(p,q))$ is given and we will show how to isometrically embed $G(I(p,q))$ in ${\mathbb Z}^2$ (the passage from the uniform to the non-uniform grid is the same as in the general case described in the proof of Proposition \ref{interval_zk}).  To embed $G(I(p,q)),$ it suffices to detect in linear time the 2-connected components of $G(I(p,q))$ or of its boundary $\partial  G(I(p,q)),$ to embed each  block $B_i$ or its boundary $\partial B_i,$ and to compose these embeddings into a single one by identifying the images of common articulation vertices of blocks (see the discussion preceding this proposition). Therefore we can suppose without loss of generality that $G(I(p,q))$ is 2-connected. Then, up to translations, symmetries, and rotations by $90^{\circ},$  $G(I(p,q))$ has a unique isometric embedding $f$ into the square grid and ${\mathcal K}(I(p,q))$ is embedded in ${\mathbb R}^2$ as a rectilinear polygon $P=P(I(p,q))$ which is monotone with respect to the coordinate axes. Each vertex of $P$ (including $f(p)$ and $f(q)$) is either a {\it convex vertex} (i.e., the interior
angle of $P$ between its two incident edges is equal to $90^{\circ}$)  or a {\it reflex vertex} (i.e., the interior angle between its two incident edges is equal to  $270^{\circ}$). Convex vertices of $P$ are exactly the images of vertices of degree 2 of $G(I(p,q))$ (in particular, $f(p)$ and $f(q)$ are convex vertices of $P$), while reflex vertices are images of vertices of degree 4 lying on the boundary of $G(I(p,q)).$  Note also that the vertices of degree 3 of $\partial G(I(p,q))$ are mapped to points  lying on sides of $P.$

Let $\pi_1$ and $\pi_2$ denote the two disjoint shortest $(p,q)$-paths constituting the boundary of $G(I(p,q)).$ To find the polygon $P(I(p,q))$ which is the image of ${\mathcal K}(I(p,q))$ under an
isometric embedding of $G(I(p,q))$ in ${\mathbb Z}^2,$ it suffices to find the images of $\pi_1$ and $\pi_2$ under such an embedding. For this, first we scan $\pi_1$ and $\pi_2$ in order
to detect the convex and the reflex vertices of each of them (note that the convex and the reflex vertices on each of these two paths alternate). Suppose that convex and reflex vertices
subdivide $\pi_1$ into the subpaths $\pi^1_1,\ldots,\pi_1^m.$ Then we define the image of $p$ to be the point $(0,0)$, draw the image of $\pi^1_1$ as a vertical path with
one end at $(0,0)$ and having length equal to the length of $\pi^1_1,$ then draw the image of the path $\pi^2_1$ as a horizontal path with the beginning at the point where the former
path ended and of length equal to the length of $\pi^2_1,$ and so on, on step $i$ we draw the image of the current path $\pi^i_1$ to be orthogonal to the image of the previous
path $\pi^{i-1}_1$ (the direction on $\pi^i_1$ depends on whether  $\pi^{i-1}_1$ and $\pi^i_1$ share a convex or a reflex vertex). To draw the images of the subpaths $\pi_2^1,\ldots,\pi_2^{m'}$ of $\pi_2$, we proceed in the same way but we start by drawing the image of $\pi^1_2$ as a horizontal path. This embedding of $\partial G(I(p,q))$ extends in a natural way to an embedding of $G(I(p,q)):$ the image of each vertex $v$ lying on the convex path (hyperplane) with ends
$u'\in \pi_1$ and $u''\in \pi_2$ is a vertex $f(u)$ of the horizontal or vertical path between $f(u')$ and $f(u'')$ and
lying on distance $d(v,u')$ from $f(u')$ and $d(v,u'')$ from $f(u'').$ Notice that $f$ is an isometric embedding of $G(I(p,q))$ into the grid ${\mathbb Z}^2$ because, up to
$90^{\circ}$'s rotations, there exists a unique isometric embedding in which $p$ is mapped to $(0,0)$ and this embedding necessarily satisfies the properties of $f$ ($f$ itself is defined in the canonical way). $\Box$
\end{proof}

\subsection{The data structure $\mathcal D$ and the computation of  $\partial G(I(p,q))$}

In this subsection, we design  the data structure $\mathcal D$ for general CAT(0) rectangular complexes, for ramified rectilinear polygons,  and for squaregraphs, allowing us quickly to compute for two arbitrary vertices $p,q$ of $\mathcal K$ the boundary paths $\pi_1$ and $\pi_2$ of $G(I(p,q))$ and the degrees $deg_0(z)$ of all vertices $z\in \pi_1\cup\pi_2=\partial(G(I(p,q)))$. The main requirement to $\mathcal D$ is the trade-off between the space occupied by $\mathcal D$ and the time for computing $\pi_1$ and $\pi_2.$ Further, we will assume that $n$ denotes the number of vertices of $\mathcal K.$ Notice that any CAT(0) rectangular complex contains $O(n)$ edges and faces. Indeed,
$|E({\mathcal K})|\le 2n$ and $|F({\mathcal K})|\le n,$ because any cube-free median graph $G$ contains a vertex $w$
of degree at most 2 (if the degree of $w$ is 2, then it belongs to a unique rectangular cell of $\mathcal K$, otherwise if the degree of $w$ is 1, then $w$ is incident to a unique edge) and, removing $w,$ the resulting graph $G'$
is also cube-free and median. As is shown in \cite{BaChEp_ramified}, as a vertex $w$ of degree at most 2 one can select any furthest vertex from a given base-point vertex. Since $G'$ contains $n-1$ vertices, $|E({\mathcal K})|-1$ or $|E({\mathcal K})|-2$ edges, and $|F({\mathcal K})|$ or $|F({\mathcal K})|-1$ faces, the required inequalities for $G$ follows by applying induction assumption to $G'.$ Finally notice that in all three cases (general CAT(0) rectangular complexes, ramified rectilinear polygons, and squaregraphs) at articulation points of $G(I(p,q))$ the order of paths $\pi_1$ and $\pi_2$ (which one is left and which one is right looking from $p$) may change.

\bigskip\noindent
{\it CAT(0) rectangular complexes.}  In case of general CAT(0)
rectangular complexes, using Breadth-First-Search (BFS), first we compute the distance matrix $D$ of  $G({\mathcal K}).$  Additionally, running BFS starting from any vertex $u$ of $V({\mathcal K}),$  for each vertex $v$ we compute the list of neighbors $L_u(v)$ of $v$ in the interval $I(u,v)$ (these are exactly the neighbors of $v$ which have been labeled by BFS before $v$). Notice that each list $L_u(v)$ contains one or two vertices.  Now, $\mathcal D$ includes the distance matrix $D$  of the underlying graph $G({\mathcal K})$ and the lists  $L_u(v), u,v\in V({\mathcal K}).$  ${\mathcal D}$ requires $O(n^2)$ space
and can be constructed in $O(|V({\mathcal K})||E({\mathcal K})|)=O(n^2)$ time.

Now, we will show how to use $\mathcal D$ to  construct the boundary paths $\pi_1$ and $\pi_2$ of $G(I(p,q))$ in $O(d(p,q))$ time. Before describing the algorithm, first notice that, when requested, the degree $deg_0(z)$ of
each vertex $z\in I(p,q)$ (and therefore of each $z\in \pi_1\cup \pi_2$) in the graph $G(I(p,q))$  can be computed in constant time by setting $deg_0(z):=|L_p(z)|+|L_q(z)|\le 4$. To build the paths $\pi_1$ and $\pi_2,$ we initialize $\pi_1:=\{ p\}=:\pi_2.$  Now suppose that after $k\ge 0$ steps, $x$ is the last vertex inserted in $\pi_1$ and $y$ is the last vertex inserted in $\pi_2$ (if $k=0$, then $x=p=y$). Notice that $x$ and $y$ have the same distance $k$ to $p$ and the same distance $d(p,q)-k$ to $q$. We will show how to compute in constant time the neighbor $x'$ of $x$ in $\pi_1$ and the neighbor $y'$ of $y$ in $\pi_2.$ From the definition of the boundary paths $\pi_1$ and $\pi_2$ follows that each of the edges  $xx'$ and $yy'$ belongs in ${\mathcal K}(I(p,q))$ to at most one rectangular cell.  To compute $x'$,  we distinguish three cases (the vertex $y'$ is computed in a similar way or is defined  together with $x'$):

\medskip\noindent
{\bf Case 1:} $x=y$.

\medskip\noindent
If $x=y$ coincides with $q,$ then the algorithm halts and returns the paths $\pi_1$ and $\pi_2.$ Otherwise, either $x=y$ coincides with $p$ or is an articulation vertex of $G(I(p,q))$. In both cases, if $L_q(x)$ consists of a single vertex $a,$ then we set $x':=a, y':=a$  and $\pi_1:=\pi_1\cup \{ a\}, \pi_2:=\pi_2\cup \{ a\}.$ On the other hand, if $L_q(x)$ consists of two different
vertices $a$ and $b,$ then set $x':=a,y':=b$ and $\pi_1:=\pi_1\cup \{ a\}, \pi_2:=\pi_2\cup \{ b\}.$

\begin{figure}\label{Case2}\centering
\begin{tabular}{ccc}
   \includegraphics[width=0.32\textwidth]{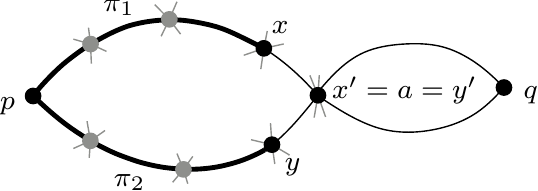}
   &
   \includegraphics[width=0.3\textwidth]{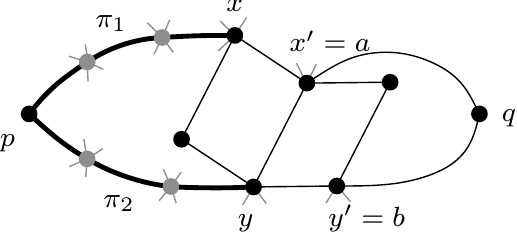}
   &
   \includegraphics[width=0.28\textwidth]{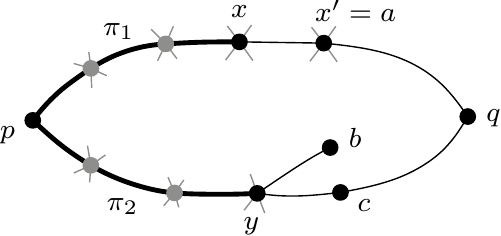}
   \\
   (a) $L_q(y)=\{a\}$
   &
   (b) $L_q(y)=\{a,b\}$
   &
   (c) $L_q(y)=\{b,c\}$
\end{tabular}
\caption{The subcases of Case 2.} \label{des_case2}
\end{figure}

\medskip
Now suppose that $x\ne y.$

\medskip\noindent
{\bf Case 2:} $|L_q(x)|=1,$ say $L_q(x)=\{ a\}.$

\medskip\noindent
Then clearly  $a$ is the next neighbor of $x$ in $\pi_1,$ therefore $x':=a$ and $\pi_1:=\pi_1\cup \{ a\}.$ Additionally, if $L_q(y)=\{ a\},$ then set $\pi_2:=\pi_2\cup \{ a\}$ and $y':=a$ (Fig. 5(a)). On the other hand, if $L_q(y)=\{ a,b\},$ then we can set $\pi_2:=\pi_2\cup \{ b\}$ and $y':=b$ (Fig. 5(b)).  Indeed, since the edge $ya$ belongs to two rectangular faces defined by $x,a,y,$ and the median of the triplet $p,x,y$ and by $y,a,b,$ and the median of the triplet $q,a,b,$ the vertex $a$ cannot be the neighbor of $y$ in $\pi_2.$   Finally, if $L_q(y)=\{ b,c\},$ then
the choice of which of the vertices $b$ and $c$ is $y'$ can be done in the same way as the choice of the vertex $x' $ in Case 3 below (Fig. 5(c)).

\medskip\noindent
{\bf Case 3:} $|L_q(x)|=2,$  say $L_q(x)=\{ a,b\}.$

\medskip\noindent
First note that since $x'$ is one of the vertices $a$ or $b,$ from Lemma \ref{L_q} below follows that the case $deg_0(a)=deg_0(b)=4$ is impossible. Suppose without loss of generality that $deg_0(a)\le 3.$ If $deg_0(b)=4,$ then Lemma \ref{L_q} implies that $x'$ cannot be the vertex $b,$  thus we can set $\pi_1:=\pi_1\cup \{ a\}$ and $x':=a$ (Fig. 6(a)) Now suppose that also $deg_0(b)\le 3$.
Then both vertices $a$ and $b$ belong to the boundary of ${\mathcal K}(I(p,q)),$ one to $\pi_1$ and another one to $\pi_2.$ Hence, $y$ is necessarily adjacent to at least one of
the vertices $a,b.$ Notice that $y$ cannot be adjacent to both $a$ and $b,$ otherwise the vertices $a,b,x,y,$ and the median of the triplet $x,y,p$ induce a forbidden $K_{2,3}$ (Fig. 7(a)). 
Therefore, to decide which of $a,b$ is $x'$ and which is $y'$ it suffices to consider to which of the two vertices $a,b$ is adjacent $y.$ If $b\in L_q(y)$ (and therefore $a\notin L_q(y)$),
then we set $\pi_1:=\pi_1\cup \{ a\},$ $x':=a$ and $\pi_2:=\pi_2\cup \{ b\},$ $y':=b$ (Fig. 6(b)). Finally,  if $a\in L_q(y)$ (and   $b\notin L_q(y)$), then we set $\pi_1:=\pi_1\cup \{ b\},$ $x':=b$ and $\pi_2:=\pi_2\cup \{ a\},$ $y':=a$ (Fig. 6(c)). 

\begin{figure}\centering
\begin{tabular}{ccc}
   \includegraphics[width=0.28\textwidth]{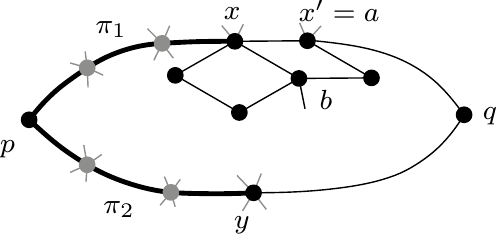}
   &
   \includegraphics[width=0.3\textwidth]{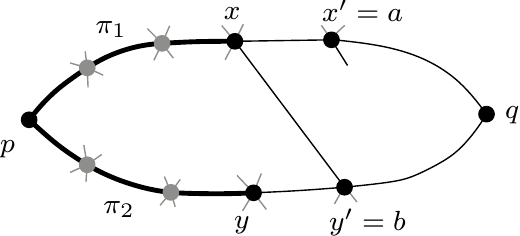}
   &
   \includegraphics[width=0.3\textwidth]{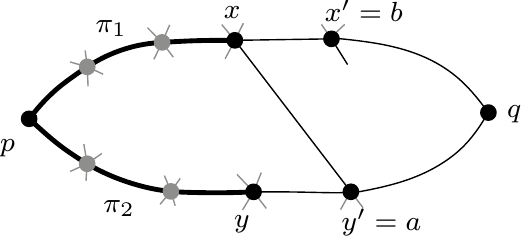}
   \\
   (a) $deg_0(b)=4$
   &
   (b) $deg_0(b)\leq 3,b\in L_q(y)$
   &
   (c) $deg_0(b)\leq 3,a\in L_q(y)$
\end{tabular}
\caption{The subcases of Case 3.} \label{des_case3}
\end{figure}


\begin{lemma}\label{L_q} If $|L_q(x)|=2$ and $x'$ is the next neighbor of $x$ in $\pi_1,$  then $deg_0(x')\le 3.$
\end{lemma}

\begin{proof} Suppose by way of contradiction that the degree $deg_0(x')$ of $x'$ in $G(I(p,q))$ is 4. Then necessarily $x'$ is either an articulation vertex or a reflex vertex. We will show  that in both cases we must have $|L_q(x)|=1,$ contrary to our assumption that $|L_q(x)|=2.$ Let $L_q(x)=\{ x',x''\}.$ Denote by $z$ the median of the triplet $x',x'',q.$ Now, if $x'$ is an articulation vertex, then necessarily $x'$ must be the closest to $q$ vertex of the 2-connected component  containing the vertex $x.$ Since the 4-cycle $C=(x,x',x'',z,x)$ also belongs to this component and $z$ is closer to $q$ than $x',$ we obtain a contradiction. Now suppose that $x'$ is a reflex vertex. This means that in ${\mathcal K}(I(p,q))$ $x'$ belongs to three rectangular cells defined by the 4-cycles $C_1=(x,u,a,x',x), C_2=(a,v,b,x',a),$ and $C_3=(b,w,c,x',b)$ of $G(I(p,q))$. The neighbors $a,x$ of $x'$ are closer to $p$ than $x'$ and the  neighbors  $b,c$ of $x'$ are closer to $q$ that $x'.$ Therefore $z$ is one of the vertices $b$ or $c.$ In both cases we conclude that the edge $xx'$ belongs in ${\mathcal K}(I(p,q))$ to two rectangular cells $C$ and $C_1,$ contrary to the assumption that $xx'$ is an edge of the boundary  path $\pi_1$  (Fig. 7(b)\& (c)). $\Box$
\end{proof}

\begin{figure}\centering
\begin{tabular}{ccc}
   \includegraphics[width=0.3\textwidth]{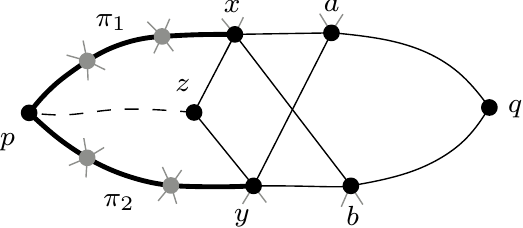}
   &
   \includegraphics[width=0.35\textwidth]{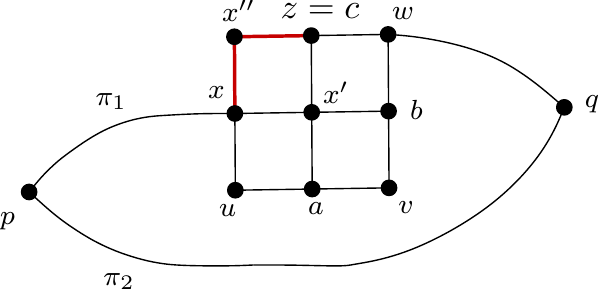}
   &
   \includegraphics[width=0.35\textwidth]{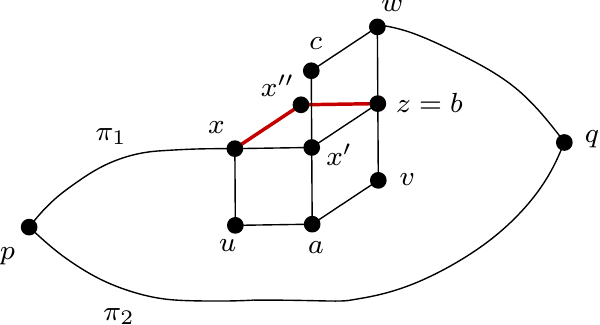}
   \\
   (a) $L_q(x)=\{a,b\}=L_q(y)$
   &
   (b) $z=c$
   &
   (c) $z=b$
\end{tabular}
\caption{To Case 2 and to the proof of  Lemma \ref{L_q}.} \label{des_lemme4.1}
\end{figure}


\bigskip\noindent
{\it Ramified rectilinear polygons.}  In case of ramified rectilinear polygons $\mathcal K$,  the data structure $\mathcal D$ consists of an isometric embedding of  $G({\mathcal K})$ into the Cartesian product of two trees $T_1,T_2$ and a data structure for nearest common ancestor queries in trees build on each of the tree-factors $T_1$ and $T_2$. As was shown in \cite{BaChEp_ramified}, the trees $T_1$ and $T_2$ and the isometric embedding of $G:=G({\mathcal K})$ in $T_1\times T_2$ can be obtained in the following way. First we construct the equivalence classes $\Theta_1,\ldots,\Theta_m$ of the Djokovi\'c-Winkler relation $\Theta$ on the edge set of $G.$ As we noticed already, $\Theta$   is the transitive closure of the ``opposite" relation
of edges of rectangular cells of $\mathcal K$, and therefore the equivalence classes of $\Theta$ can be easily constructed in total
$O(|F(\mathcal K)|+|E(\mathcal K)|+|V(\mathcal K)|)=O(n)$ time. Simultaneously with $\Theta$, we can construct the {\it incompatibility graph} Inc$(G)$ of $G$:
the equivalence classes $\Theta_1,\ldots,\Theta_m$ are the vertices of Inc$(G)$ and  two equivalence classes $\Theta_i$ and $\Theta_j$ define an edge  of Inc$(G)$ if and only if there exists a rectangular face of $\mathcal K$ in which two opposite edges belong to $\Theta_i$ and two other opposite edges belong to $\Theta_j$ (the definition of Inc$(G)$ is different but equivalent to that given in \cite{BaChEp_ramified}).  Clearly, Inc$(G)$ can be constructed in linear time: it suffices to consider each of the rectangular faces of $\mathcal K$ and to define an edge between the two equivalence classes sharing edges with this face. The graph Inc$(G)$ is bipartite \cite{BaChEp_ramified}, moreover a coloring of vertices of Inc$(G)$ (i.e., of equivalence classes of $\Theta$) in two colors defines the two tree-factors $T_1$ and $T_2.$ (As is shown in Proposition 1 of \cite{BaChEp_ramified} this is equivalent to a coloring of edges of $G$ in two colors such that opposite edges of each 4-cycle $C$ have the same color and incident edges of $C$ have different colors.)

The trees $T_i$ $(i=1,2)$ are obtained
from $G$ by collapsing all edges colored with a color different from $i.$ Equivalently, to construct $T_i,$ we remove the edges (but keep their ends) colored with color $i$ and compute the connected components of the  resulting graph $G_i.$ Each connected component $C$ of $G_i$ defines a vertex of $T_i$ and two connected components $C',C''$ of $G_i$ define an edge $C'C''$ of $T_i$  if there exists an edge of $G$ (necessarily colored $i$) with one end in $C'$ and another end in $C''.$ Notice that each edge of $T_1$ or $T_2$ is labeled by a different equivalence class of $\Theta$.  Now, if $u$ is an arbitrary vertex of $G$ and $C_1(u)$ and $C_2(u)$ are the connected components of $G_1$ and $G_2,$ respectively, containing the vertex $u,$ then the couple $f(u)=(C_1(u),C_2(u))$ are the coordinates of $u$ in the isometric embedding $f$ of $G$ in $T_1\times T_2.$ The trees $T_1,T_2$ and the coordinates of the vertices of $G$ in $T_1\times T_2$ can be defined in total $O(n)$ time. The trees $T_1$ and $T_2$ are preprocessed in linear time to answer in constant time lowest common ancestor queries  \cite{HaTa}.  Additionally, for each vertex $v$ we define the sorted list $Q(v)$ of the equivalence classes of $\Theta$ to which belong the edges incident to $v.$ These lists $Q(v), v\in V(\mathcal K),$ occupy total linear space. These are the three constituents  of the data structure $\mathcal D.$

Now, we will show how to use $\mathcal D$ to construct the boundary paths $\pi_1$ and $\pi_2$ of $G(I(p,q))$ and the degrees $deg_0(z)$ in $G(I(p,q))$ of their vertices $z$ (for an illustration, see Fig. 8). First, given the coordinates $f(p)=(C_1(p),C_2(p))$ and $f(q)=(C_1(q),C_2(q))$ of $p$ and $q$ in $T_1\times T_2,$ using two lowest common ancestor queries, one for $C_1(p)$ and $C_1(q)$ in $T_1$ and the second one for $C_2(p)$ and $C_2(q)$ in $T_2,$ we can compute the path $P_1$ connecting $C_1(p)$ and $C_1(q)$ in $T_1$ and the path $P_2$ connecting $C_2(p)$ and $C_2(q)$ in $T_2$ in time proportional to the number of edges on these paths.  Then $G(I(p,q))$ can be isometrically embedded in the Cartesian product $P_1\times P_2.$ We will call $P_1$ and $P_2$  vertical and horizontal paths, respectively.  We also suppose that
$f(p)$ and $f(q)$ are respectively the lowest leftmost and the upper rightmost corners of $P_1\times P_2.$

\begin{figure}\centering
\includegraphics[width=0.9\textwidth]{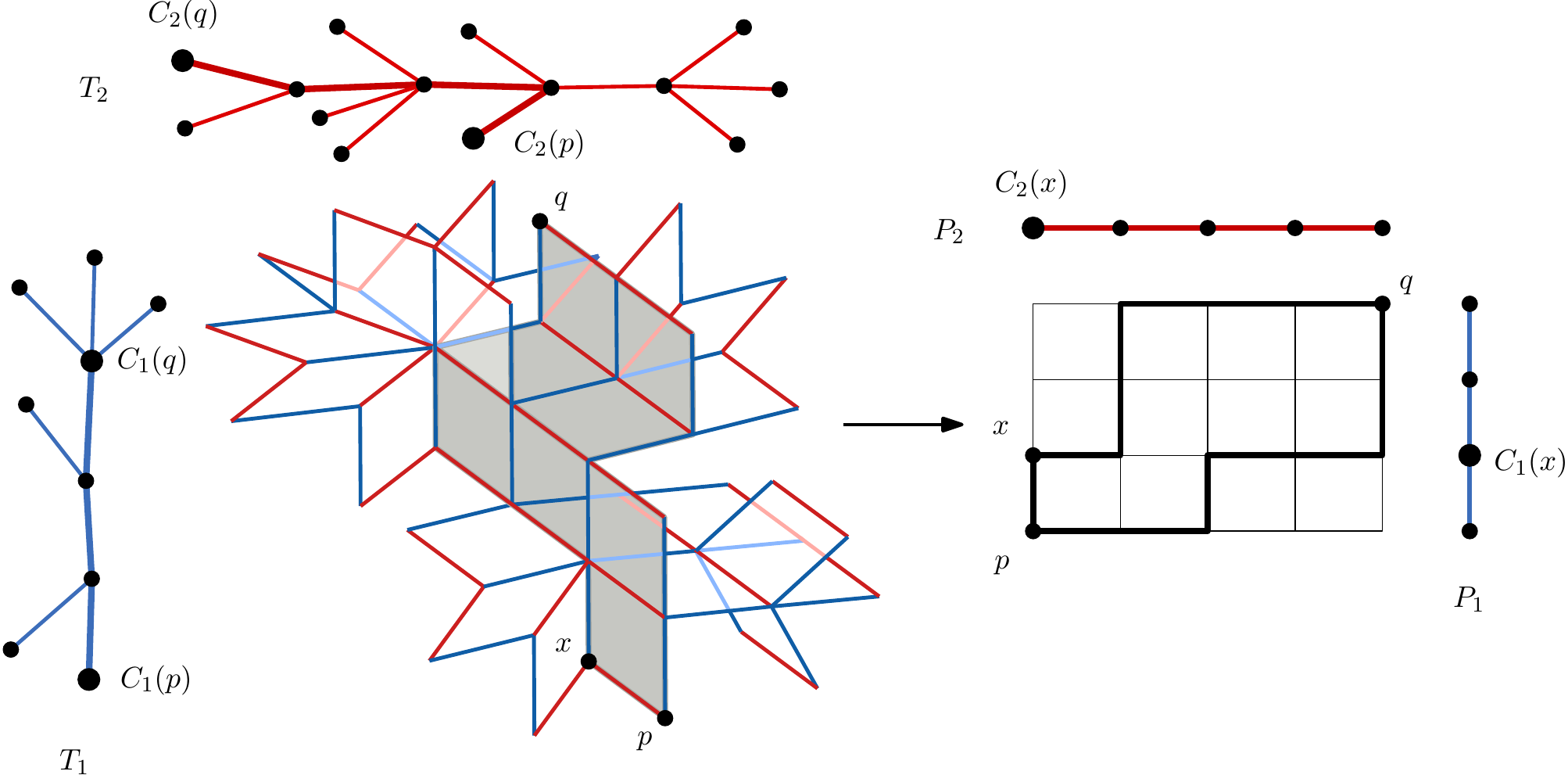}
\caption{Construction of $\pi_1$ and $\pi_2$ in a ramified rectilinear polygon.} \label{pi_ram-des}
\end{figure}

We start by setting $\pi_1:=\{ p\}=:\pi_2$ and we will construct $\pi_1$ and $\pi_2$ in such a way that $\pi_1$ is the upper path and $\pi_2$ is the lower path of the embedding of $G(I(p,q))$ in $P_1\times P_2.$ Let $x$ be the last vertex of $\pi_1$ and we will show how to define the next vertex $x'$ of $\pi_1.$ Let $f(x)=(C_1(x),C_2(x)).$ Suppose that  the next edge  of $P_1$ incident
to $C_1(x)$ is labeled by $\Theta_i$ and the next edge of $P_2$ incident to $C_2(x)$ is labeled by $\Theta_j.$ Using binary search on the sorted list $Q(x)$ we can decide in $O(\log (deg(x)))$ time if $x$ has an incident edge belonging to the equivalence class $\Theta_i$ and/or an incident edge belonging to $\Theta_j.$ If $x$ has an incident edge $e$ from $\Theta_i,$ then we set $x'$ to be the end-vertex of $e$ which is different from $x$ (notice that $C_2(x')=C_2(x)$). In this case, the path $\pi_1$ goes vertically. Otherwise, if no edge of $\Theta_i$ is incident to $x,$ then necessarily there is an edge $e'$ of $\Theta_j$ incident to $x$ and, in this case, we set $x'$ to be the end-vertex of $e'$ which is different from $x.$ Then $C_1(x')=C_1(x)$ and the path $\pi_1$ goes horizontally.  Analogously, if $y$ is the last vertex of $\pi_2,$ in order to  define the next vertex $y'$ of $\pi_2,$ we consider the labels $\Theta_i$ and $\Theta_j$ of the next edges of the paths $P_1$ and $P_2,$ respectively, and using binary search on $Q(y)$ we decide if $y$ has an incident edge belonging to $\Theta_j$ and/or an incident edge belonging to $\Theta_i.$ If $y$ has an incident edge $e$ from $\Theta_j,$ then we set $y'$ to be the other end-vertex of $e$ and in this case the path $\pi_2$ goes horizontally. Otherwise, if no edge of $\Theta_j$ is incident to $y,$ then  there is an edge $e'$ of $\Theta_i$ incident to $y$ and in this case we set $y'$ to be the end-vertex of $e'$ different from $y$ and the path $P_2$ goes vertically.

Notice also that in a similar way we can compute in $O(\log (deg(z)))$ time the degree $deg_0(z)$ in $G(I(p,q))$ of each vertex $z\in \pi_1\cup \pi_2.$ As a consequence, the paths $\pi_1,\pi_2$ and the degrees $deg_0(z), z\in \pi_1\cup \pi_2$ can be computed in total $O(d(p,q)\log\Delta)$ time, where $\Delta$ is the maximum degree of a vertex of $G(\mathcal K).$

\bigskip\noindent
{\it Squaregraphs.} Finally, in the case of squaregraphs,  as a data structure $\mathcal D$ we will take an encoding of  plane graphs defining polygonal complexes of nonpositive curvature presented in \cite{ChDrVa_jalg}. Let $G=G(\mathcal K).$ The data structure from \cite{ChDrVa_jalg} uses vertex labels of size $O(\log^2 n)$ bits (for each vertex of $G$ it uses a label consisting of $O(\log n)$ integers of length at most $\log n$) and allows for each pair $u,v$ of vertices of $G$ to compute in constant time the distance $d(u,v)$ between $u$ and $v$ in $G$ and a neighbor $u'$ of $u$ lying on a shortest path between $u$ and $v.$ Additionally, we suppose that the plane graph $G$ is represented as a planar subdivision in the form of a doubly-connected edge list \cite{BeChKrOv}. Each edge of $G$ belongs to one or two rectangular faces and, using this representation, the vertices and the edges belonging to these faces can be listed in constant time.

Now, given two vertices $p$ and $q,$ in order to construct the boundary paths $\pi_1$ and $\pi_2$ of $G(I(p,q))$ and the degrees $deg_0(z)$ of the vertices $z\in \pi_1\cup \pi_2$ in $G(I(p,q))$  we proceed in the following way. The algorithm of \cite{ChDrVa_jalg} returns in $O(1)$ time a neighbor $p'$ of $p$ in the interval $I(p,q).$ Necessarily, $p'$ is  a vertex of $\partial G(I(p,q)).$ Without loss of generality, we assign $p'$ to the path $\pi_1.$ To find the neighbor $p''$ of $p$ in $\pi_2$ we consider the edges incident to $p$ in  one or two faces containing  the edge $pp'.$ For each of the end-vertices of these edges which is different from $p$ we compute in constant time its distance to $q$ using the algorithm from \cite{ChDrVa_jalg}. If one of these vertices is closer to $q$ than $p,$ then we denote it by $p''$ and insert it in $\pi_2.$ Otherwise, we set $p'':=p'.$

Now suppose that $x$ is the last vertex of $\pi_1$ and that $x_0$ is the vertex preceding $x$ in $\pi_1.$ Suppose also that $deg_0(x_0)$ and the neighbors  of $x_0$ in $G(I(p,q))$ have been already computed. We will show now how to compute in constant time $deg_0(x),$ the neighbors of $x$ in $G(I(p,q)),$ and the next neighbor $x'$ of $x$ in $\pi_1$ (the computation of $deg_0(y),$ the neighbors of $y$ in $G(I(p,q)),$ and the next neighbor $y'$ of $y$ in $\pi_2$ can be done in a similar way). For this, we consider the rectangular faces of $\mathcal K$ incident to the edge $x_0x$ (there are at most two such faces) and performing in constant time distance queries \cite{ChDrVa_jalg} to $q$, we compute which vertex $x_1$ (if it exists) adjacent to $x$ from these faces belongs to $I(p,q)$ (since the edge $x_0x$ belongs to $\pi_1,$ at most one such vertex can belong to $I(p,q)$). Again we distinguish several cases.

First suppose that the vertex $x_1$ exists and $x_1\in I(p,q)$. Then we consider the face of $\mathcal K$ (if it exists) incident to $xx_1$ and not containing $x_0$ and test if the second neighbor $x_2$ of $x$ in this face belongs to $I(p,q).$ If $x_2$ belongs to
$I(p,q),$ then again we test if the second face incident to the edge $xx_2$ exists and if the neighbor $x_3$ of $x$ in this face belongs to $I(p,q).$ If $x_3$ exists and $x_3\in I(p,q),$ then we set
$x':=x_3$ and $deg_0(x):=4$ because in this case we have $L_p(x)=\{x_0,x_1\}$ and $L_q(x)=\{ x_2,x_3\}.$ On the other hand, if $x_2$ exists and $x_2\in I(p,q)$ but $x_3$ does not exist or $x_3\notin I(p,q),$ then we set $x':=x_2$ and $deg_0(x):=3$ (in this case, $x_0,x_1,x_2$ are the three neighbors of $x$ in $I(p,q)$). Otherwise, if $x_2$ does not exist or $x_2\notin I(p,q),$ then we set $deg_0:=2$ and $x':=x_1$ if $x_1\in L_q(x).$ Finally, if $x_1\in L_p(x),$ then $x$ is an articulation vertex of $G(I(p,q))$ and we proceed $x$ in the same way as the vertex $p.$ Namely, we set $x'$ to be the neighbor of $x$ in $I(x,p)$ returned by the algorithm of \cite{ChDrVa_jalg} (as we noticed already, this vertex necessarily belongs to $\partial G(I(p,q))$).   To compute $deg_0(x)$ and the neighbors of $x$ in $I(p,q)$ we need to find if $x$ has another vertex in $L_q(x).$ For this, we consider the edges incident to $x$ in  the faces incident to the edge $xx'.$ For each of the end-vertices of these edges which is different from $x'$ we compute its distance to $q$ using the algorithm from \cite{ChDrVa_jalg}. If one of these vertices is closer to $q$ than $x,$ then we include it in $L_q(x)$ and set $deg_0(x):=4,$ otherwise we set $deg_0(x):=3.$

Now suppose that  $x_1$ does not exist or $x_1$ does not belong to $I(p,q).$ Then $x$ is an articulation vertex of $G(I(p,q)).$ Again, we set $x'$ to be the neighbor of $x$ in $I(x,q)$ returned by the algorithm of \cite{ChDrVa_jalg} and check if $x$ has another vertex in $L_q(x).$ If such a neighbor exists, then we include it in $L_q(x)$ and return $deg_0(x):=3,$ otherwise we return $deg_0(x):=2$ (in this case, $L_p(x)=\{ x_0\}$ and $L_q(x)=\{ x'\}$).  For a given $x,$ each of the operations used to compute $x'$ and $deg_0(x)$ requires constant time, therefore, using the data structure $\mathcal D$ of size $O(n\log n)$ we can construct the boundary path $\pi_1$ and compute the degrees in $G(I(p,q))$ of its vertices in total $O(d(p,q))$ time. The path $\pi_2$ can be constructed in a similar way and within the same time bounds.

\subsection{The algorithm}

Summarizing the results of the previous subsections, we are ready to present the main steps of the algorithm for answering shortest path
queries in CAT(0) rectangular complexes,  ramified rectilinear polygons, and squaregraphs.

\bigskip
\begin{center}
\framebox{
\parbox{15cm}{
\vspace{0.05cm}
\noindent{\bf Algorithm} {\sc Two-point shortest path queries}\\
{\footnotesize
  {\bf Input:} A CAT(0) rectangular complex ${\mathcal K},$ a data structure $\mathcal D$, and two points $x,y\in {\mathcal K}$\\
  {\bf Output:} The shortest path $\gamma(x,y)$ between $x$ and $y$ in ${\mathcal K}$\\
  \begin{tabular}[t!]{l@{ }p{14cm}}
  1. & Given the rectangular faces containing the points $x$ and $y,$ compute the vertices $p,q$ of ${\mathcal K}$ such that $x,y\in {\mathcal K}(I(p,q)).$\\
  2. & Using the data structure ${\mathcal D},$ compute the boundary $\partial G(I(p,q))$ and the degrees $deg_0(z)$ in $G(I(p,q))$ of the vertices $z$ of $\partial G(I(p,q))$.\\
  3. & Using the algorithm described in the proof of Proposition \ref{interval_z2}, compute an unfolding $f$ of $\partial G(I(p,q)).$ Let $P(I(p,q))$ denote the chain of monotone polygons bounded
  by $f(\partial G(I(p,q))).$\\
  4. & Locate $f(x)$ and $f(y)$ in  $P(I(p,q))$.\\
  5. & Using the algorithm for triangulating monotone polygons (see, for example, \cite{BeChKrOv}), triangulate each monotone polygon constituting a block of $P(I(p,q)).$\\
  6. & In the triangulated polygon $P(I(p,q))$ run the algorithm of Lee and Preparata \cite{LeePre} and return the shortest path $\gamma^*(f(x),f(y))=(f(x),z_1,\ldots,z_m,f(y))$
  between $f(x)$ and $f(y)$ in  $P(I(p,q)),$ where
  $z_1,\ldots,z_m$ are all vertices of $P(I(p,q)).$\\
  7. & Return $(x,f^{-1}(z_1),\ldots,f^{-1}(z_m),y)$ as the shortest path $\gamma(x,y)$ between the points $x$ and $y.$
  \end{tabular}}
\\ }}
\hspace{0.2cm}
\end{center}
\medskip

It remains to specify how to  implement the steps 1 and 2 of the algorithm. For step 1, given two points $x$ and $y$ of $\mathcal K,$ we are also given two rectangular faces $R(x)$ and $R(y)$ containing $x$ and $y.$ Then using the distance matrix $D$ for CAT(0) rectangular complexes, the coordinates of the embedding in the case of ramified rectilinear polygons, and the distance queries from \cite{ChDrVa_jalg} for squaregraphs, we can compute two furthest vertices $p$ and $q,$ where $p$ is a vertex of $R(x)$ and $q$ is a vertex of $R(y).$ This takes constant time because we take the maximum of a list of 16 distances between the vertices of $R(x)$ and $R(y).$ In the case of ramified rectilinear polygons, to compute the distance $d(u,v)$ in constant time, it suffices in each tree $T_i$ $(i=1,2)$  to keep in $\mathcal D$ the distance $d_{T_i}(C,R_i)$ in $T_i$ from each vertex $C$ to the root $R_i$ of $T_i$. Now, if $f(u)=(C_1(u),C_2(u))$ and $f(v)=(C_1(v),C_2(v)),$ then we compute the lowest common ancestor $C_1$ of $C_1(u)$ and $C_1(v)$ in the tree $T_1$, the lowest common ancestor $C_2$ of $C_2(u)$ and $C_2(v)$ in $T_2,$ and return as $d(u,v)$ the value $(d_{T_1}(C_1(u),R_1)+d_{T_1}(C_1(v),R_1)-2d_{T_1}(C_1,R_1))+(d_{T_2}(C_2(u),R_2)+d_{T_2}(C_2(v),R_2)-2d_{T_2}(C_2,R_2)).$ Notice also that the step 4 of the algorithm also requires constant time: having the coordinates of $x$ in $R(x)$ and of $y$ in $R(y),$ since $R(x)$ is the unique rectangular face incident to $p$ in ${\mathcal K}(I(p,q))$ and $R(y)$ is the unique face incident to $q$, we can easily locate the images of $R(x)$ and $R(y)$ in the polygon $P(I(p,q)).$ Summarizing, here is the main result of this paper:

\begin{theorem} \label{two-point_queries} Given a CAT(0) rectangular complex $\mathcal K$ with $n$ vertices, one can construct a data structure $\mathcal D$ of size $O(n^2)$ so that, given any two points $x,y\in \mathcal K,$ we can compute the shortest $l_2$-path $\gamma (x,y)$ between $x$ and $y$ in $O(d(p,q))$ time, where $p$ and $q$ are  vertices of two faces of $\mathcal K$ containing the points $x$ and $y,$ respectively, such that $\gamma(x,y)\subset {\mathcal K}(I(p,q))$ and $d(p,q)$ is the distance between $p$ and $q$ in the graph $G(\mathcal K).$ If $\mathcal K$ is a ramified rectilinear polygon, then one can construct a data structure $\mathcal D$ of optimal size $O(n)$ and answer two-point shortest path queries in $O(d(p,q)\log\Delta)$ time, where $\Delta$ is the maximal degree of a vertex of $G(\mathcal K).$ Finally, if $\mathcal K$ is a squaregraph, then one can construct a data structure $\mathcal D$ of  size $O(n\log n)$ and answer two-point shortest path queries in $O(d(p,q))$ time. \end{theorem}

\bigskip\noindent
{\bf Open questions:} (1) We do not know how to design a subquadratic data structure $\mathcal D$ allowing to perform two-point shortest path queries in  CAT(0) rectangular complexes in $O(d(p,q))$ time or how to use the encoding provided by the isometric embedding of ramified rectilinear polygons into products of two trees to remove the logarithmic factor in the query time.

\medskip\noindent
(2) It will be interesting to generalize our algorithmic results (using Propositions 1-3) to all  CAT(0) box complexes, in particular to 3-dimensional  CAT(0) box complexes.

\section*{Acknowledgement}

\noindent
We wish to thank the anonymous referees
for a careful reading of the first version of the manuscript and useful suggestions.
This work was supported in part by the ANR grants  OPTICOMB (ANR BLAN06-1-138894) and GGAA.

\end{document}